\documentclass[traditabstract]{aa} 
\usepackage{graphicx}
\usepackage{txfonts}
\usepackage{supertabular}
\usepackage{textcomp}
\usepackage{booktabs}
\usepackage{lscape}
\usepackage{natbib}
\usepackage{xspace}
\usepackage{tabularx}
\graphicspath{{./}{images/}}
\usepackage{txfonts}
\begin{document}
	\title{Megamaser detection and nuclear obscuration in Seyfert galaxies}

	\author{Michael Ramolla
		\inst{1}
		\and
		Martin Haas
		\inst{1}
		\and
                Vardha Nicola Bennert
		\inst{2}
		\and
		Rolf Chini
		\inst{1,3}
		}

	\institute{Astronomisches Institut, Ruhr--Universit\"at Bochum,
		Universit\"atsstra{\ss}e 150
                , 44801 Bochum, Germany 
		\and
                Department of Physics, University of California,
                Santa Barbara, CA 93106, USA
                \and 
                Facultad de Ciencias, Universidad Cat\'{o}lica del Norte, Antofagasta, Chile\\}

	\date{Received June 21, 2010; accepted February 10, 2011}

 
	\abstract{
          We revisit the relation between H$_{\rm 2}$O maser detection rate and nuclear obscuration for a sample of 114 Seyfert galaxies, drawn from the CfA, 12$\mu$m and IRAS F25/F60 catalogs. These sources have mid-infrared spectra from the Spitzer Space Telescope and they are searched for X-ray and [O\,III]$\,{5007\AA}$ fluxes from the literature. We use the strength of the [O\,IV]\,25.9$\mu$m emission line as tracer for the intrinsic AGN strength. After normalization by [O\,IV] the observed X-ray flux provides information about X-ray absorption. The distribution of X-ray~/~[O\,IV] flux ratios is significantly different for masers and non-masers: The maser detected Seyfert-2s (Sy 1.8-2.0) populate a distinct X-ray~/~[O\,IV] range which is, on average, about a factor four lower than the range of Seyfert-2 non-masers and about a factor of ten lower than the range of Seyfert-1s (Sy 1.0-1.5). Non-masers are almost equally distributed over the entire X-ray~/~[O\,IV] range. This provides evidence that high nuclear obscuration plays a crucial role for the probability of maser detection. Furthermore, after normalization with [O\,IV], we find a similar but weaker trend for the distribution of the maser detection rate with the absorption of the 7\,$\mu$m dust continuum. This suggests that the obscuration of the 7 $\mu$m continuum occurs on larger spatial scales than that of the X-rays. Hence, in the AGN unified model, at moderate deviation from edge-on, the 7\,$\mu$m dust absorption may occur without proportionate X-ray absorption. The absorption of [O\,III] appears unrelated to maser detections. The failure to detect masers in obscured AGN is most likely due to insufficient observational sensitivity.  
	}

	\keywords{
          Galaxies: Seyfert -- 
          Galaxies: nuclei -- 
          Masers -- 
          X-rays: galaxies --
          Infrared: galaxies
	}
		\maketitle
%

\section{Introduction}
	
H$_{\rm 2}$O megamaser galaxies represent an extreme subclass of
active galactic nuclei (AGN) with strong water maser emission at 
22~GHz (reviews by \citealt{2005ARA&A..43..625L} and
\citealt{2005A&A...436...75H}). 
In those cases where the emission arises from a molecular disk and can
be resolved spatially using Very Long Baseline 
Interferometry, the central black hole (BH) mass and the distance to
the galaxy can be determined
(e.g. for NGC\,4258, \citealt{1993ApJ...406..482G},
\citealt{1999Natur.400..539H}).  
Thus, finding megamasers {\it (henceforth simply called masers)}  
and understanding their properties is of great interest. 

From theoretical considerations, a large line-of-sight column density
of velocity coherent gas favors the detection of a maser.  
High velocity coherence of the maser emitting gas is required, because
energy and momentum conservation imply that the induced photon has the
same frequency and direction as the stimulating photon
(e.g. \citealt{2002IAUS..206..452E}). 
While the emission of an individual maser spot is directional 
(i.e. beamed), a collection of such spots statistically may be
expected to radiate in all directions, but this has not been confirmed
so far. 
The originally discovered water maser emission from AGN comes from
(presumably edge-on) disks, and the resolved emission in most sources 
traces accretion disks while a few cases are star formation masers. 
However, two sources, Circinus and NGC 3079, show in addition
also off-disk jet masers that seem to trace outflows. 
Potentially 
these outflow masers are actually torus clouds
(\citealt{2008ApJ...685..147N}).  

In the AGN unified model, an
optically thick obscuring dust torus is envisioned to encircle the
accretion disk and type-1 AGN are seen pole-on while type-2 AGN are
seen edge-on (\citealt{1993ARA&A..31..473A}). 
Masers are almost exclusively found in AGN of Seyfert-2 or LINER
type, consistent with the picture that masers are preferentially 
beamed in the plane of the torus 
(\citealt{1997ApJS..110..321B,2004ApJ...617L..29B},
\citealt{2005A&A...436...75H}).  
But not all type-2 AGN are masers.
{ 

From the conceptional viewpoint, it should be noted that the 22GHz radio-frequency maser
emission itself is believed to be largely unaffected by absorption; but a high X-ray, optical or 
mid-infrared obscuration may signpost a high likelihood that the masing disk is seen edge-on,
hence favoring a maser detection. 
}

Type-2 AGN that host masers show a prevalence ($>80\%$) 
of high X-ray obscuring columns ($N_{\rm H} > 10^{23} \rm{cm}^{2}$) 
and about half are Compton thick ($N_{\rm H} > 10^{24} \rm{cm}^{2}$) 
(\citealt{1997ApJS..110..321B}, \citealt{2006A&A...450..933Z},
\citealt{2008ApJ...686L..13G}).  
However, as pointed out by \citet{2006A&A...450..933Z}, among type-2
AGN 
the average X-ray derived column densities of masers and
non-masers\footnote{
We denote as non-masers those AGN that have been observed at
22~GHz, but for which no megamaser was detected.
} 
are indistinguishable. 
One explanation for this unexpected result could be that X-ray 
scattering in clumpy media dilutes the true 
line-of-sight column density, and thus prevents us from deriving 
unbiased orientation information. 
{ Therefore it is vital to include also information from other than X-ray wavelengths, 
to reveal the potential influence of nuclear obscuration on the maser detection and non-detection,
respectively. }

Recently, \citet{2010ApJ...708.1528Z} analyzed the 
K$\alpha$ iron line equivalent width EW(K$\alpha$), 
following the strategy of \citet{1999ApJS..121..473B}, and compared it
with two optical thickness parameters, 
the infrared 6-400\,$\mu$m luminosity L$_{\rm IR}$ derived from IRAS
12-100\,$\mu$m photometry and 
the [O\,III]\,5007\AA~ emission line luminosity L$_{\rm [O\,III]}$. 
Both parameters were  
adopted to be isotropic tracers for the intrinsic AGN strength. 
While the EW(K$\alpha$) distributions of 19 masers and 34 non-masers
cover the same broad range (100 - 3000 eV),  
the median EW(K$\alpha$) of masers is about a factor 4 higher 
than that of the non-masers, indicating that the X-ray continuum of 
masers is more absorbed than that of non-masers. 

{ However it is still a matter of debate, whether L$_{\rm IR}$ and L$_{\rm [O\,III]}$
are indeed isotropic tracers of the intrinsic AGN luminosity.
While [O\,III] has often been used as isotropic AGN tracer 
(\citealt{1994ApJ...436..586M,1997MNRAS.288..977A,1999ApJS..121..473B,2005ApJ...634..161H,2006A&A...455..173P,2009A&A...504...73L}),
the discovery of polarized [O\,III] emission in some type-2 AGN (\citealt{1997A&A...328..510D})
cautions that a substantial fraction of [O\,III] can be shielded by the torus.
Further studies, using MIR emission lines like [O\,IV] or [Ne\,V] as orientation independent tracer
of the AGN power, provide evidence by means of the [O\,III]/[O\,IV] ratio,
that [O\,III] suffers
}  orientation dependent extinction, up to a
factor of 10 in individual cases (\citealt{2005A&A...442L..39H}, 
\citealt{2008ApJ...682...94M}, \citealt{2010ApJ...710..289B}). 

{ 

This is qualitatively consistent with results obtained using the (extinction corrected) 2-10keV X-ray 
luminosity L$_{\rm X}$ as intrinsic AGN power measure; \citet{2006A&A...453..525N} find that 
L$_{\rm [O\,III]}$\,/\,L$_{\rm X}$ 
of type-2 AGN is, on average, about a factor two lower than that of type-1 AGN. 
From a conceptional viewpoint, even}
in the face-on Sy1 case, the back-sided cone of
the NLR lies -- at least partly -- behind an absorbing layer (e.g. the
dust torus). 
{ Therefore it is highly questionable how far [O\,III] can serve as an isotropic AGN tracer.}
The
extinction correction via Balmer 
decrement (H$_\alpha$~/~H$_\beta$ = 3) remains highly uncertain, since
it is dependent on the geometry of the emitting and obscuring regions.

\citet{1996A&A...313..423H,1997MNRAS.286...23B} caution against the use of [O\,III] as
a measure of the intrinsic NLR emission and
suggest to use [OII] 3727 instead.
Observations of radio-loud AGN, where the orientation can be inferred
from radio morphology, show that [O\,II] is largely orientation independent 
(\citealt{1996A&A...313..423H,1997MNRAS.286...23B}). On the other hand, 
because of its low ionization
potential,
[O\,II] can also be dominated by star formation in the host (e.g. \citealt{2005ApJ...629..680H}).
Then, the decline of L$_{\rm [O\,II]}$\,/\,L$_{\rm X}$ with increasing L$_{\rm X}$, 
as found by \citet{2006A&A...453..525N},
could be naturally explained by a decline of host\,/\,AGN with increasing AGN L$_{\rm X}$.

Likewise the mid-infrared ($\lambda < 40 \mu$m) part of L$_{\rm IR}$ is 
orientation dependent (e.g. Fig. 16 in \citealt{2006AJ....132..401B}), 
while the far-infrared ($\lambda > 40 \mu$m) emission of Seyfert
galaxies and low-luminosity quasars actually is dominated by star 
forming contributions rather than by the AGN itself
(e.g. \citealt{1995ApJ...454...95M}, \citealt{2006ApJ...649...79S}). 
Thus, a careful re-investigation using more suited isotropic AGN
tracers would be desirable.  
 
Here, we revisit the connection between maser detection rate and nuclear
obscuration using the strength of the [O\,IV] 25.9$\mu$m emission line 
(for short [O\,IV]) as tracer for the intrinsic AGN strength.
[O\,IV] has been found to be largely
unaffected by obscuration (e.g., \citealt{1998ApJ...498..579G}, 
\citealt{2005A&A...442L..39H}, \citealt{2008ApJ...682...94M},
\citealt{2010ApJ...710..289B}).  
We combine the strategies of \citet{1999ApJS..121..473B} and
\citet{2008ApJ...682...94M}. 
The observed X-ray (2-10 keV) flux normalized by [O\,IV] should 
provide information about X-ray absorption, even in the case of X-ray
scattering caused by a complex geometry or for Compton thick cases. 
We compare the distribution of X-ray~/~[O\,IV] for masers
and non-masers. 
In addition, after normalization with [O\,IV], 
we inspect the relation between 
maser detection rate and absorption of the 7\,$\mu$m dust continuum
emitted from the nuclear torus, 
as well as maser detection and the absorption of the 
[O\,III]\,5007\AA~ emission of the central part of the 
narrow-line-region (NLR). 
  
The distances from which we derived the luminosities are taken from
the NED database. The cosmology is based on
H$_{\rm o}=73$~km~s$^{-1}$~Mpc$^{-1}$, $\Omega_{\Lambda} =0.73$ and
$\Omega_{m} = 0.27$.

\section{Data}

\subsection{The parent sample}
{
At first glance, one could take all known masers and non-masers from the literature
and compare their properties, for instance L$_{\rm X}$\,/\,L$_{\rm [O\,IV]}$. But in order to 
determine nuclear obscuration, one needs to know also the range of L$_{\rm X}$\,/\,L$_{\rm [O\,IV]}$
for unobscured (preferentially Sy1) sources, which should comprise a complete sample free
from any selection bias. However, the list of Seyferts, for which a maser 
search has been performed, did not follow clear selection criteria. 
Even worse, most maser searches have been performed on Sy2s, but only a small number on Sy1s.
Because incomplete sample selection may influence the results, we here decided to start with 
complete Seyfert catalogs having well defined selection criteria. 
In order to increase the sample size, we created a 
master sample from the following three catalogs,
} consisting 
of { a total of 163} sources. 

\begin{itemize}
\item [$\bullet$] The magnitude limited complete sample of the CfA
  Redshift Survey by 
  \citet{1992ApJ...393...90H}, which was supplied with 
  updated Seyfert-type information from the NED 
  database. 
\item [$\bullet$] The 12$\mu$m Active Galaxy Sample by
  \citet{1989ApJ...342...83S}, 
  complemented by \citet{1993BAAS...25Q1362R}.  
\item [$\bullet$] The IRAS F$25$/F$60$ flux-ratio selected sample by
  \citet{1992A&AS...96..389D}, as refined by
  \citet{2003ApJS..148..327S}. 
\end{itemize}
{ 
Table 1 documents how the 163 sources distribute over the three catalogs,
and how these catalogs match or complement each other. In general, we will
present the results for the combined sample, but -- where necessary -- also
for the catalogs individually (Tab. \ref{samplenumber_table}).

Further below (Section \ref{sample_properties}) we will discuss potential differences between
the three samples and 
our combined sample and all other known masers 
outside of it.
}
The Spitzer data archive contains IRS spectra (at $\sim 26\mu$m) for
{  126 of the 163} 
sources
classified as Seyferts according to the NED.
{ This data is listed in Tab. \ref{table1}.}
It covers the complete CfA-sample 
of 54 Seyfert Galaxies.
It includes { 107} of 118 Seyferts
(two Blazars included as Sy1) from the
12$\mu$m selected sample. 
For the IRAS sample 
we found useful IRS spectra for 34 of 60 sources. 

\subsection{Maser information}\label{Maser}

The parent sample of { 126} 
 sources with Spitzer spectra 
was searched for known maser-detections and non-detections. 
For this purpose
we used the lists as 
compiled by \citet{2009ApJ...695..276B} and on the website of 
the Hubble Constant Maser  
Experiment (HoME)\footnote{\tiny
  http://www.cfa.harvard.edu/$\sim$lincoln/demo/HoME/surveys/survey.html
  compiled from 
  \citet{2006ApJ...649..561K,2006ApJ...652..136K,2003MNRAS.344L..53H,2006A&A...450..933Z,2004ApJ...617L..29B,2002A&A...383...65H,2002ApJ...565..836G,1997ApJ...486L..15G,2005PASJ...57..587S,1996ApJS..106...51B,1995A&A...304...21G,1995PASJ...47..771N,1998A&A...335..463H,2005A&A...436...75H,2003ApJS..146..249B,1984A&A...141L...1H,1985Natur.314..144H,1986ApJ...308..592C,1986A&A...155..193H,1993A&A...268..483B,1990ApJ...364..513G,2008ApJ...678...96B,2008Henkel,2008dde..confE...5B} 
}.

{
This search results in 18 masers (3 Sy1s, 15 Sy2s), {96 } non-masers ({36} Sy1s, 60 Sy2s) and  
{12} sources ({10} Sy1s, 2 Sy2s) for which no maser search has been performed so far (henceforth called 
maser-unknown). The results are listed in Tab. \ref{table1}, Col. 2.
}
\subsection{[O\,IV] 25.89\,$\mu$m  Line  and $7\,\mu$m Continuum Flux}

Our analysis is based on public archival IRS spectra of Seyfert
 galaxies. 
We used the 
post-basic-calibration data (PBCD), as reduced by the Spitzer Science
Center's (SSC) pipeline. 
This included droop-, stray-light-, cross-talk- and
saturation correction, dark subtraction, flatfielding and coaddition.

If possible, the IRS high resolution spectra with R $\sim 600$ have been
chosen, to avoid contamination of [O\,IV]$_{25.89\,\mu m}$ with 
the neighboring [Fe\,II]$_{25.99\,\mu m}$ emission line. 
If high-resolution spectra were not available, the low-resolution 
spectra were used, including a background subtraction, which  
was also performed by the SSC pipeline. 

For the high resolution data, collected with the shorter (4.7 x 11.3
and 11.1 x 22.3 arcsec) slits, separate background observations had to
be chosen to evaluate the background contribution.
This was performed in \citet{Ramolla:Thesis:2009}, { by comparing 
the background with the source fluxes at the presumably weakest part of the 
source spectrum between $9$ and $10¨\mu$m rest frame;}
with the result that the background contribution is negligible in
comparison with the conservatively 
assumed flux calibration errors of $15\%$.
The resulting errors are calculated from an assumed $15\%$ flux
calibration error and the error of the line fitting routine.

{ The [O\,IV] flux has been extracted by fitting a simple spectral model 
in a wavelength-window of $\sim 0.3~\mu$m around the [O\,IV] line. This 
model consists 
of a linear base, convolved with Gaussian profiles that also include 
the neighboring [Fe\,II] line.}
No [O\,IV] aperture corrections had to be applied, because 
for both, high- and low-resolution data, the slit apertures cover
an area larger than the expected size of the NLR, as estimated 
from the relationship\footnote{ $log({\rm R}_{\rm NLR}) = (0.52 \pm 0.06)\,\times\, log({\rm L}_{\mbox{[O\,III]}}) - (18.5 \pm 2.6)$} 
of [O\,III] luminosity to NLR size, found by \citet{2002ApJ...574L.105B}. 

We calculated the $7.6\,\mu$m (henceforth for short $7\,\mu$m) 
continuum flux from 
the background-subtracted IRS low-resolution spectra. 
We used a modified version of the PAHFIT code by
\citet{2007ApJ...656..770S} which estimated the continuum in the $5$
to $11.8~ \mu$m branch. As suggested by
\citet{pahfit-descr:2008}, we did not correct the continuum fit for the
silicate feature at $9.7\,\mu$m. The $7\,\mu$m continuum flux is then
calculated from the PAH flux and the equivalent-width of the
features at $7.4~\mu$m, $7.6~\mu$m and $7.8~\mu$m 
\citep{Ramolla:Thesis:2009}.  
The uncertainties of the $7\,\mu$m continuum are conservatively 
estimated to be smaller than 30\%, which is sufficient for our purpose. 
In a few cases the AGN contribution may be contaminated 
by nuclear ($<$3.7$\arcsec$) star formation 
(e.g. \citealt{2009ApJ...705...14D}). We checked that the effect on our statistical 
analysis is negligible by comparison with high resolution ground-based MIR
observations.

The [O\,IV] and $7\,\mu$m fluxes are listed in Tab. \ref{table1}. 
The values are consistent with those derived by others 
(\citealt{2006AJ....132..401B}, 
\citealt{2007ApJ...671..124D}, \citealt{2009ApJ...705...14D}, 
\citealt{2010ApJ...709.1257T}).

\subsection{X-rays + [O\,III] 5007\,\AA~line from the literature}
\label{X-rays}

 The 2-10\,keV hard X-Ray data have been obtained by several observers
 using ASCA, Beppo SAX,
 Chandra and XMM. 
We collected the data from the NED; in case of
multiple entries we chose the latest detection. 

We have collected [O\,III]\,5007\AA~ emission line fluxes 
from { various literature sources, as listed in Tab. \ref{table1}}. 
Because of the large uncertainties, we did neither apply
any extinction correction nor any aperture
correction for the [O\,III] fluxes. 
Such aperture corrections would affect  
a few very nearby sources, but most sources are sufficiently distant
so that in the statistical analysis any bias is small.

\subsection{Additional maser sources}
{
On the one hand, our combined sample is drawn from the CfA, 12$\mu$m and IRAS F25/F60 catalogs,
containing 15 Sy2 maser sources with Spitzer spectra. 
On the other hand a total of 52 masing Sy2 are known so far 
(\citealt{2009ApJ...695..276B,2010ApJ...708.1528Z}), although drawn from different
AGN catalogs using inhomogeneous criteria.

In order to compare the 15 Sy2 maser of our combined sample with the remaining
37, we also analyzed available Spitzer spectra and gathered further 
[O\,III] and X-ray fluxes for them from the literature in the same manner,
as we did on our combined sample. This results in an ``off-sample'' list
of 37 Sy2 masers that is appended to Tab. \ref{table1}.
}
\section{Results and discussion}

{ While our combined sample of 126 sources contains {114} maser and non-maser 
sources ({12} maser-unknown excluded)}, 
not all of them have data in all observables considered here 
(X-rays, F$_{\rm 7 \mu m}$, [O\,III]). 
Therefore, we compare the maximum possible subsets for pairs of
observables, and discuss the 
implications in the framework of the AGN unified model. 
Therein we consider as components the accretion disk, supposed to house
the maser, the 
dust torus, the bi-conical NLR and the host galaxy. 
We here denote by Sy1s the subtypes between Seyfert 1.0 and 1.5, 
and by Sy2s those between Seyfert 1.8 and 2.0.
{ 
All Figures contain an combined error bar in the lower right corner that is 
averaged from all relative errors in this measure. Since the literature 
sources did not uniformly presented errors, we do not perform this step for the
X-ray, [O\,III] and H$_2$O luminosities.}

\subsection{Nuclear X-ray obscuration}
\label{section_nuclear_obscuration}

\begin{figure}
	\includegraphics[angle=0,width=\columnwidth]{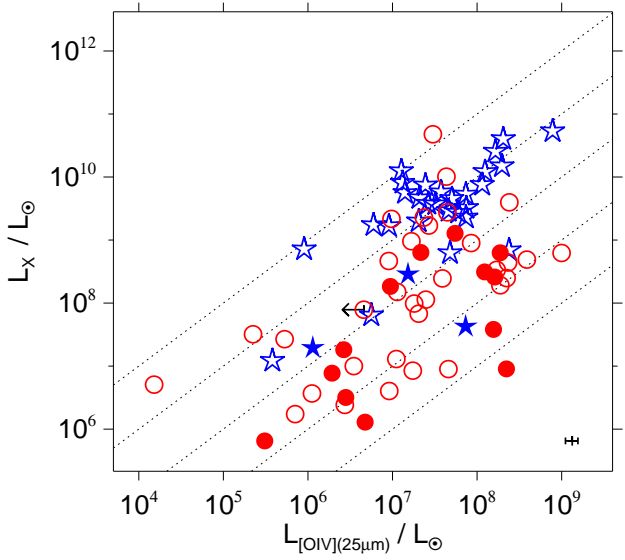}
	\caption{
          Observed 2-10 keV X-ray versus [O\,IV] line luminosity 
          Blue stars represent Sy1s (Sy 1.0-1.5), 
          red circles Sy2s (Sy 1.8-2.0). 
          Filled symbols are masers, open symbols are non-masers. 
	  The dotted lines mark fixed L$_{\rm X}$\,/\,L$_{\rm
          [O\,IV]}$ ratios of 
	  1000; 100; 10; 1; 0.1 (from top to bottom).
          { The error-bar in the lower 
            right corner is the average relative error of all [O\,IV]
          measurements.}
        }
	\label{fig_o4_xo}
\end{figure}

\begin{figure}
	\includegraphics[angle=0,width=\columnwidth]{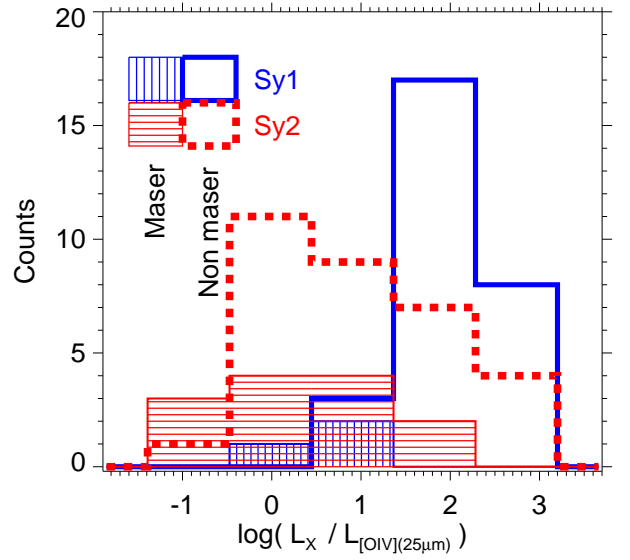}
	\caption{
	Histogram of the L$_{\rm X}$\,/\,L$_{\rm
          [O\,IV]}$ ratio of the data
        points shown in Fig. \ref{fig_o4_xo}. 
        The red dashed line 
	represents Sy2 non-masers, whereas the blue thick 
	line represents the Sy1 non-masers. 
        The maser-detections are 
	represented by the dashed surfaces - blue and vertically 
        dashed for Sy1, red and horizontally dashed for Sy2. 
        { The one upper limit is excluded.}
        }
	\label{Xo-o4_LUM_msr-HIST}
\end{figure}

Figure~\ref{fig_o4_xo} shows the observed hard X-ray versus [O\,IV]
line luminosities, and Figure~\ref{Xo-o4_LUM_msr-HIST} the 
L$_{\rm X}$\,/\,L$_{\rm [O\,IV]}$ { histogram}.
The 
features are:
\begin{itemize}

\item [$\bullet$]
 { On average, Sy1s present an about 10 times higher X-ray\,/\,[O\,IV]
    ratio than Sy2s.}
\item [$\bullet$] 
Sy2 non-masers are evenly distributed over the entire range 
occupied by Sy2 masers and by Sy1s (Fig.~\ref{fig_o4_xo}). 

\item [$\bullet$]
Sy2 masers are almost disjoint from Sy1s. At a given [O\,IV]
luminosity, Sy2 masers have on average about a factor { 4} 
lower X-ray
luminosity than Sy2 non-masers (Fig.~\ref{Xo-o4_LUM_msr-HIST}). 
Likewise, the few Sy1 masers have a lower X-ray
luminosity than Sy1 non-masers.\footnote{ 
The Sy1 masers are NGC\,4051, NGC\,4151 and UCG\,5101. Note that both 
NGC\,4051 and NGC\,4151 have relatively low maser H$_{\rm 2}$O
luminosity, and UCG\,5101 is an ultra-luminous infrared galaxy so that 
the maser luminosity could arise from starburst regions rather than
from the AGN accretion disk.
}
\item [$\bullet$]
 { {6} out of {12} sources without masing information, but 
L$_{\rm X}$ and L$_{\rm [O\,IV]}$ available (see Tab. \ref{table1}), show the same trends as the Sy1s and
Sy2s with masing information (see Tab. \ref{samplenumber_table}). They are not plotted, 
to avoid overcrowding of Figs. \ref{fig_o4_xo} and \ref{Xo-o4_LUM_msr-HIST} with too many different symbols.
}

\end{itemize}
We assume that the X-ray deficit, i.e. the decrease of 
L$_{\rm X}$\,/\,L$_{\rm [O\,IV]}$, is caused by obscuration, probably in
the molecular dust torus. 
Then, the Figures~\ref{fig_o4_xo} and \ref{Xo-o4_LUM_msr-HIST}  
clearly demonstrate that masers are
found almost exclusively in Sy2s with heavy nuclear obscuration, 
while non-maser Sy2s exhibit a broad range of X-ray absorption. 
A two-sided Kolmogorov-Smirnov (KS) test, applied on the L$_{\rm X}$\,/\,L$_{\rm [O\,IV]}$ distribution, 
results in a probability of
{ $18\%$ }
that the Sy2 masers and non-masers are drawn from the same
parent population. 
Our results agree with those of 
\citet{2008ApJ...686L..13G} and \citet{2010ApJ...708.1528Z} who
find that about $60\%$  
of the masers are Compton-thick.

\begin{figure}
	\includegraphics[angle=0,width=\columnwidth]{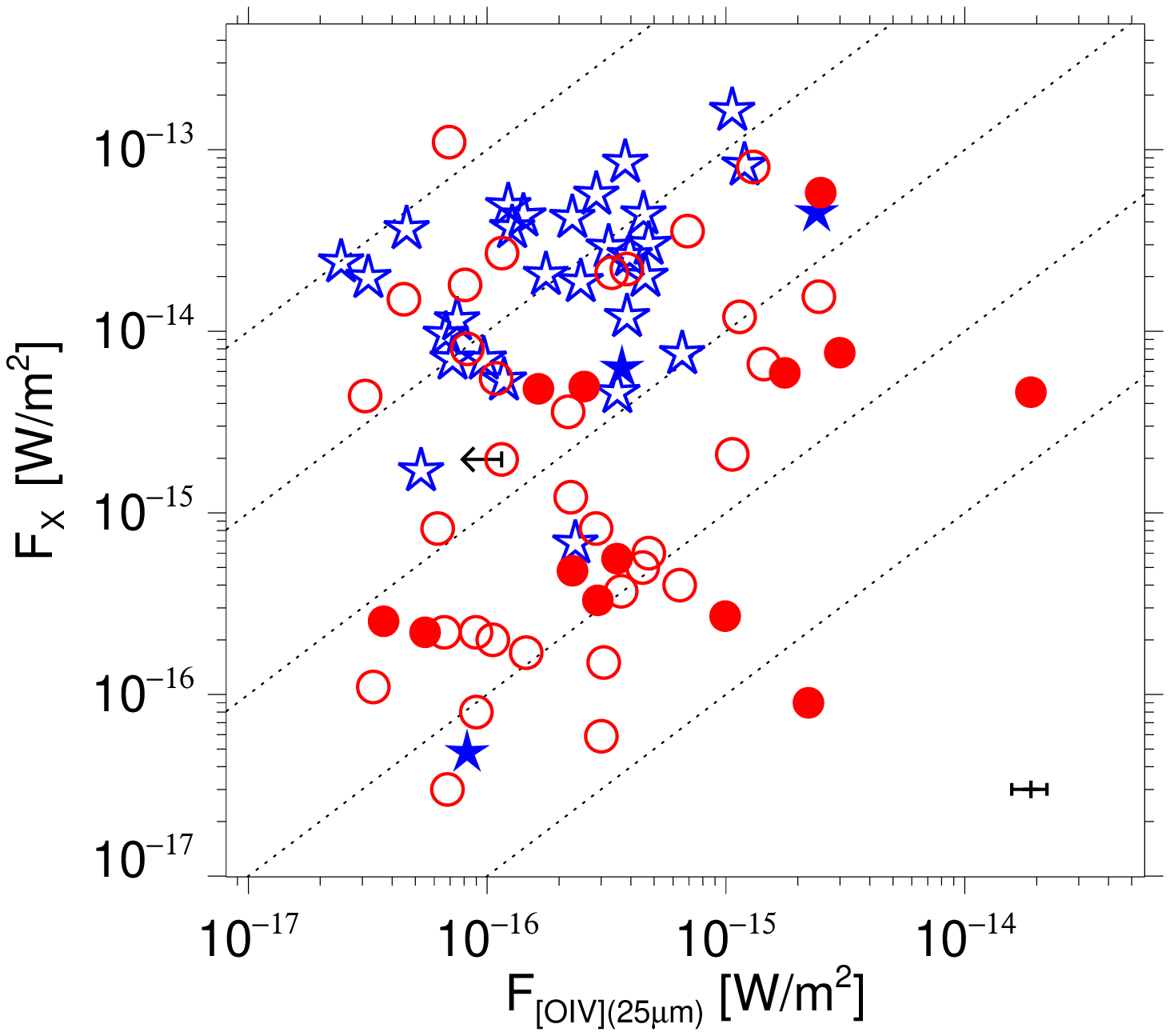}
	\caption{
          Observed X-ray versus [O\,IV] line flux. 
          Symbols and colors are as in Fig. \ref{fig_o4_xo}. 
	  The dotted lines mark fixed flux ratios of
	  1000; 100; 10; 1; 0.1 (from top to bottom).
        }
 {          The error-bar in the lower 
          right corner is the average relative error of all [O\,IV]
          measurements.}
	\label{flux_o4_xo}
\end{figure}

\begin{figure}
	\includegraphics[angle=0,width=\columnwidth]{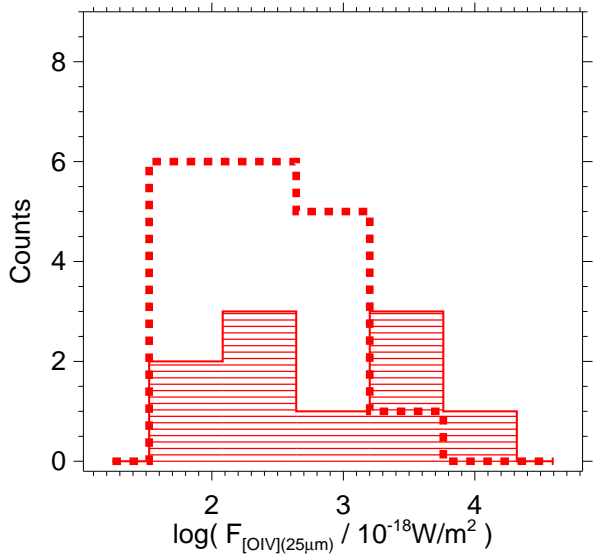}
	\caption{[O\,IV] flux histogram of Sy2 masers for absorbed
          sources from Fig.\,\ref{flux_o4_xo} with 
          L$_{\rm X}$ / L$_{\rm [O\,IV]}$ $<$ 10. 
          Masers are represented by the shaded area, non-masers by the
          thick dashed histogram. 
        }
	\label{Xo-o4_LUM_fx-sep_msr-HIST}
\end{figure}

Assuming that Sy1s are almost unobscured, the obscured sources populate
the L$_{\rm X}$\,/\,L$_{\rm [O\,IV]}$ range below 10 in
Fig~\ref{fig_o4_xo}. 
Thus, masers populate almost completely the range of obscured
sources. Surprisingly this range also contains numerous
non-masers.  
In order to better understand why in such absorbed sources the maser
search failed, we consider the influence of observed brightness. 
Fig. \ref{flux_o4_xo} shows the X-ray and [O\,IV] flux distribution 
(instead of the luminosity distribution). In fact, the Sy2 masers and
non-masers show a flux-dependence in their [O\,IV] distribution. 
Sources with
low [O\,IV] flux 
are more frequently classified as
non-masers { (2 Sy2 maser and 12 Sy2 non-maser at L$_{\rm [O\,IV]} <$ 10$^{\rm -16}$ Wm$^{\rm -2}$)}, 
while sources with
high [O\,IV] flux 
are more frequently classified as
masers {(6 Sy2 maser and 5 Sy2 non-maser at L$_{\rm [O\,IV]}>$ 10$^{\rm -15}$ Wm$^{\rm -2}$)}. 
Fig. \ref{Xo-o4_LUM_fx-sep_msr-HIST} displays the [O\,IV]
 fluxes of all Sy2s lying below the dividing line between obscured and
 unobscured sources 
(L$_{\rm X}$ : L$_{\rm [O\,IV]}$ $\approx$ 10). 
Among this subset of obscured Sy2s, the frequency of non-masers rises
constantly towards lower [O\,IV] fluxes, in contrast to the
distribution of masers. This incidence is consistent with an
observational bias against the maser-detection for faint AGN.
This implies a relation between [O\,IV] flux and H$_{\rm 2}$O flux which 
is indeed observed in Fig. \ref{o4-maser-msr_LUM}
{ and discussed in Section \ref{section_luminosity}}.
 
Because the detection of maser emission appears to be biased 
against sources with low flux,  
we conclude that among obscured sources the true fraction of masers
is higher than indicated by Fig. \ref{Xo-o4_LUM_msr-HIST}.

\subsection{Extended obscuration of the dust torus and the NLR}

\begin{figure}
	\includegraphics[angle=0,width=\columnwidth]{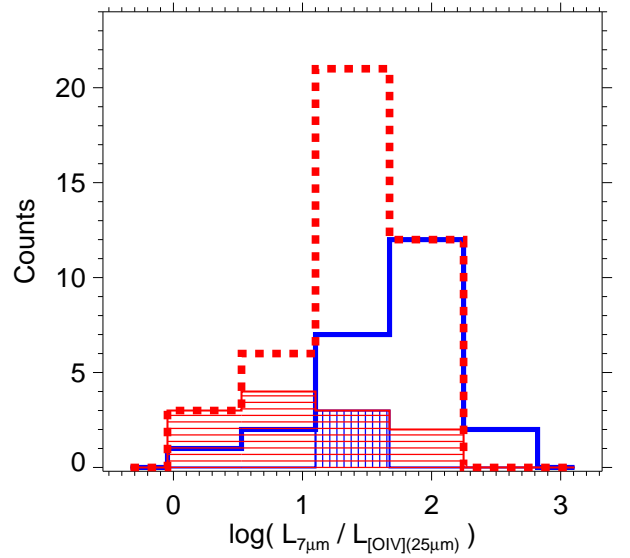}	
	\caption{
		Histogram of the $7\, \mu$m continuum to [O\,IV] line 
                ratio. 
                Legend as in Fig. \ref{Xo-o4_LUM_msr-HIST}. 
{ 
                   All 21 upper limits from Tab. \ref{table1} are excluded.}
		}
	\label{7mu-o4_LUM_msr-HIST}
\end{figure}

Figure 
\ref{7mu-o4_LUM_msr-HIST} shows { a histogram of the MIR 7\,$\mu$m continuum to }
[O\,IV] line ratio. 
The striking features of this diagram are:
\begin{itemize}
\item [$\bullet$] 
Sy2s populate about the same total range as Sy1s, but show a
prevalence for
lower 7\,$\mu$m~/~[O\,IV] values, i.e. a 7\,$\mu$m continuum deficit. 
On average, the ratio 7\,$\mu$m~/~[O\,IV] of Sy2s is about a
factor of 3 
lower than that of Sy1s.  
This is consistent with the results obtained via radio normalization
(7\,$\mu$m~/~8GHz) by Buchanan et al. (2006, their Fig. 16).  

\item [$\bullet$] 
Among Sy2s the 7\,$\mu$m~/~[O\,IV] ratio of masers is, 
on average, about a factor of {  2} 
lower than that of non-masers. 
A KS test results in a probability of { 3.7\%} 
that the
Sy2 masers and non-masers are drawn from the same parent
distribution. 
Flux considerations similar to those for L$_{\rm X}$ /
L$_{\rm [O\,IV]}$ suggest that the
true 7\,$\mu$m~/~[O\,IV] separation of masers and non-masers will be 
{ even more pronounced} once the observational bias against the detection of low 
flux masers is taken into account. 
\end{itemize}
We assume that the deficit of the 7\,$\mu$m continuum in Sy2s is mainly
caused 
by absorption of the torus dust emission. This absorption has to 
take place somewhere between the emitting region and the observer, 
hence probably in the "halo" of the torus, i.e. in the outer part of the
torus itself or in the host galaxy. 
It is possible that the scale height of this MIR-absorbing halo, 
i.e. the projected distance of absorbing material 
from the line-of-sight to the nuclear accretion disk,  
is (much) larger than the scale height of the torus itself. 
This is consistent with the results from a Spitzer study of CfA
Seyferts (\citealt{2007ApJ...671..124D}), where  
sources with high $10~\mu$m silicate absorption show a preference
for large host-galaxy inclinations and irregularities (merger events
or interactions), both of which lead to absorption through the host. 

\begin{figure}
	\includegraphics[angle=0,width=\columnwidth]{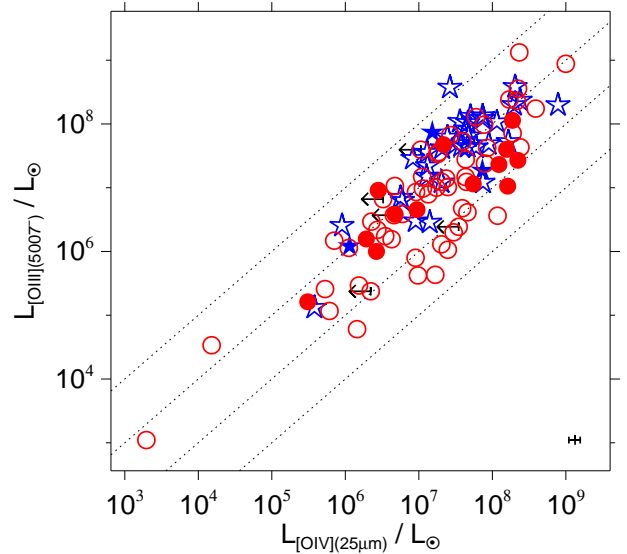}
	\caption{
          Distribution of [O\,III] 5007\AA~ versus [O\,IV] 25.9$\mu$m luminosity.
          Symbols are as in Figure~\ref{fig_o4_xo}. 
          The dotted lines mark fixed-ratios 
	  10; 1; 0.1; 0.01 
          (from top to bottom). 
{           The error-bar in the lower 
          right corner is the average relative error of all [O\,IV]
          measurements.}
        }
	\label{fig_o4_o3}
\end{figure}

\begin{figure}
	\includegraphics[angle=0,width=\columnwidth]{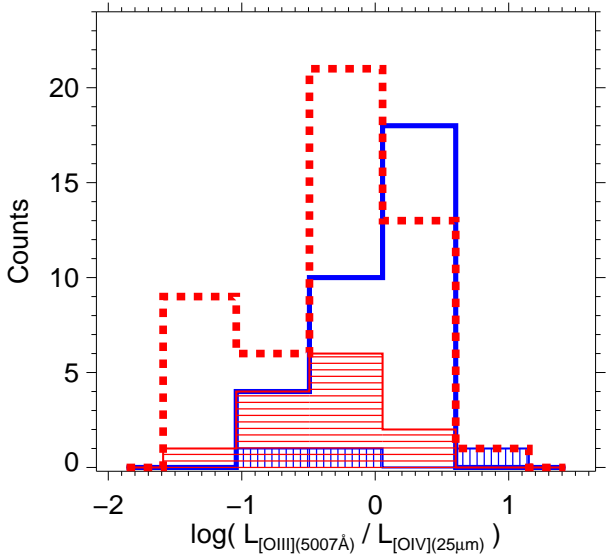}
	\caption{
          Histogram of the [O\,III] to [O\,IV] ratio. 
          Legend as in Fig. \ref{Xo-o4_LUM_msr-HIST}. { Five upper limits are excluded.}
		}
	\label{o3-o4_LUM_msr-HIST}
\end{figure}

In order to provide further clues on the extent of the MIR-absorbing
material, we consider the [O\,III] 5007\AA~ versus 
[O\,IV] line luminosity as shown in Figures \ref{fig_o4_o3} and 
\ref{o3-o4_LUM_msr-HIST}. 
The features of the [O\,III]~/~[O\,IV] distribution are 
similar to those of 7\,$\mu$m~/~[O\,IV]. 
Most Sy2s populate the same range as Sy1s, a few Sy2s
show a [O\,III] deficit, i.e. on average about a factor 3 
lower [O\,III]~/~[O\,IV] ratios when compared to Sy1s, 
consistent with
results by \citet{2010ApJ...710..289B} on the 12\,$\mu$m sample.  
The distribution pattern of masers and non-masers
appears to be statistically indistinguishable. 
A KS-test results in a probability of { 61\%}
that the Sy2 masers and non-masers are drawn from the same parent
population. 
However, the Sy2 subsample in Figures \ref{fig_o4_o3} and 
\ref{o3-o4_LUM_msr-HIST} shows a distinct tail towards lower ratios, potentially caused by absorption.

The [O\,IV] 25.9$\mu$m line is $\sim$50 times 
less affected by extinction than the optical [O\,III] $5007\AA$ line.
A low [O\,III]~/~[O\,IV] ratio argues in
favor of large obscuration, as explained in
\citet{2005A&A...442L..39H}. Another explanation of deviating
[O\,III]~/~[O\,IV] ratios could be different radiation fields in the
NLR. 
Because the [O\,IV] 25.9 $\mu$m 
line needs a higher ionization potential than the optical 
[O\,III] 5007\AA~ line, AGN with a hard radiation field are expected to
show a low ratio.
The Sy2s with low [O\,III]~/~[O\,IV] would then be those 
AGN with hard radiation fields. But this
is not consistent with other spectroscopic MIR tracers 
like the [Ne\,II] ${12.8\,\mu m}$ to
[O\,IV] flux ratio  (\citealt{2008ApJ...689...95M}). 
Thus we conclude that in Figures \ref{fig_o4_o3} and
\ref{o3-o4_LUM_msr-HIST} the NLR of Sy2s with [O\,III] deficit is 
considerably obscured. 

While the [O\,III] obscuration may occur mainly in the innermost part
of the NLR, the large extent ($>$1 kpc) of the NLR suggests 
that the absorption is not confined to the region encircled by the dust
torus. Rather the sky-projected distribution of the absorbing material 
might reach further out to a considerable distance (several hundred
parsec) from the line-of-sight to the nuclear accretion disk. 
The presence of moderately extended [O\,III] absorption, as well as
the similarity of the 7\,$\mu$m~/~[O\,IV]  and  [O\,III]~/~[O\,IV] diagrams
supports the picture that also the MIR absorption takes place in a 
moderately extended layer, i.e. the torus halo mentioned above. 
Although both observables 7\,$\mu$m and [O\,III] appear 
to be affected by absorption in a similar fashion, we note that 
the distribution of
7\,$\mu$m~/~[O\,III] spans a large range (2-700). 
This is not surprising 
in view of the diversity of the
orientation-dependent appearance of the involved emitters and
absorbers even for a simple
AGN model.  

\subsection{Combined picture}
Why do masers and non-masers show so different distributions in 
X-ray~/~[O\,IV], while their distribution in 
7\,$\mu$m~/~[O\,IV] looks more similar?  

Because masers need a large line-of-sight column density
of velocity coherent gas, 
they
are expected to be predominantly detected in edge-on accretion disks.
Thus, the maser detection or non-detection can tell us about the disk 
orientation with respect to the line-of-sight.  
In order to constrain the implications in the framework of the 
AGN unified model, we consider two extreme cases: 
\begin{enumerate}
\item 
  For a disk seen edge-on, obviously the maser is most easily detected 
  and the molecular torus is seen more or less edge-on, too. 
  In this case the nuclear accretion disk (and its corona) is shielded by the
  torus, so that the X-rays are heavily obscured. 
 If { additional} extended material, able to obscure the 
  MIR emission, does not lie in the torus plane, the 
  7\,$\mu$m~/~[O\,IV] ratio is decreased. 
 \item
For a non-maser, both, 
disk and torus appear to be sufficiently tilted away from edge-on,
so that the nuclear X-ray absorption is relatively low.
In addition, our diagrams indicate the existence of non-masers,
where the torus plane is seen edge-on,
but the disk could be tilted out of this plane due to 
locally different angular momentum.
In this case of a non-maser, the edge-on torus causes
a high obscuration of the X-ray nucleus as well. 
On the other hand, irrespective of the disk and torus orientation, the
MIR continuum can be absorbed or not depending on the line-of-sight
through the extended host.
\end{enumerate}
From these two extreme cases we see: 
While the requirement for heavy absorption of the nuclear X-rays is that 
the line-of-sight has to hit a rather compact area with very 
high column density, the area of the torus emission
and even more the area of the (bright) NLR emission is orders of 
magnitude larger, so that the absorber must cover a larger area, too.
If the absorption of X-rays and 7\,$\mu$m occurs on different spatial
scales, the strength of the obscuration in each wavelength range 
may be sensitive to small differences in the aspect angle. 
Furthermore, because the MIR-absorbing material is located 
farther away from the line-of-sight to the nucleus, it is less reliable to predict whether a maser will be detected.  


\begin{table*}[h!]
\begin{center}
  \begin{tabular}{l | c | c | c | c | c | c}
\hline
\multicolumn{7}{c}{}  \\
\multicolumn{1}{c}{(1)} &\multicolumn{1}{c}{(2)} &\multicolumn{1}{c}{(3)} &\multicolumn{1}{c}{(4)} &\multicolumn{1}{c}{(5)} &\multicolumn{1}{c}{(6)} &\multicolumn{1}{c}{(7)} \\
\multicolumn{7}{c}{}  \\
& & & & & & \\
Intersecting sample &  CfA & 12$\mu$m & IRAS & 12$\mu$m $\bigcup$ IRAS& CfA $\bigcup$ IRAS  & CfA $\bigcup$  12$\mu$m  \\
& & & & & & \\
\hline
& & & & & & \\

CfA          &$ 54  $&$ 42  $&$ 12 $&$ 45 $&$ -  $&$ - $ \\
12$\mu$m     &$ -  $ &$ 118 $&$ 24 $&$ -  $&$ 57$ &$ - $  \\
IRAS F25/F60 &$ -   $&$  - $ &$ 60 $&$ -  $&$ - $ &$ 27$  \\
\multicolumn{7}{c}{}  \\
\hline
\multicolumn{7}{c}{}  \\
  \end{tabular}
  \caption{{ 
Documentation of how far the three samples (of 163 sources in total) match and complement each other. For
each row the Table entries list the number of sources contained in the intersection of the sample denoted by 
the columns 2-7 with the CfA, 12$\mu$m and IRAS F25/F60 sample (Col. 1).
{E. g. the union of the 12$\mu$m and IRAS F25/F60 samples (Col. 5) has 45 sources in common with the CfA sample (Row 1).}
}
  }
  \label{complementation_table}
  \end{center}
\end{table*}

\begin{table*}[h!]
\begin{center}
  \begin{tabular}{r | l | c c | c c | c c | c}
\hline
\multicolumn{2}{c}{}  & \multicolumn{2}{c}{} & \multicolumn{2}{c}{} & \multicolumn{2}{c}{} & \multicolumn{1}{c}{}  \\
\multicolumn{2}{c}{(1)}  & \multicolumn{2}{c}{(2)} & \multicolumn{2}{c}{(3)} & \multicolumn{2}{c}{(4)}& \multicolumn{1}{c}{(5)} \\
\multicolumn{2}{c}{}  & \multicolumn{2}{c}{} & \multicolumn{2}{c}{} & \multicolumn{2}{c}{}& \multicolumn{1}{c}{} \\
\multicolumn{2}{c}{}  & \multicolumn{2}{c}{L$_{\rm{X}}$ / L$_{\rm{[O\,IV]}}$} & \multicolumn{2}{c}{ L$_{7\mu\rm{m}}$ / L$_{\rm{[O\,IV]}}$} & \multicolumn{2}{c}{ L$_{\rm{[O\,III]}}$ / L$_{\rm{[O\,IV]}}$} & \multicolumn{1}{c}{ Parent Sample} \\
\multicolumn{1}{c}{}& & & & & & & & \\
Parent Sample & Subset               & Number   & log(Ratio)    & Number   & log(Ratio)    & Number   & log(Ratio) & Number\\
\multicolumn{1}{c}{}& & & & & & & & \\
\hline 
\multicolumn{1}{c}{}& & & & & & & & \\
CfA        & Sy1 maser     &$ 3      $&$ 0.8 \pm 0.9 $&$ 3      $&$ 1.6 \pm 0.1 $&$ 3      $&$ 0.0 \pm 0.6 $& 3 \\
           & Sy1 non-maser &$ 16     $&$ 2.0 \pm 0.5 $&$ 14     $&$ 1.9 \pm 0.5 $&$ 16     $&$ 0.1 \pm 0.4 $& 18\\
           & Sy1 unknown   &$ 2      $&$ 1.73\pm 0.02$&$ 3      $&$ 2.1 \pm 0.6 $&$ 2      $&$ -0.1\pm 0.1 $& 4 \\
           & Sy2 maser     &$ 4      $&$ 0.1 \pm 0.5 $&$ 4      $&$ 1.2 \pm 0.7 $&$ 4      $&$ -0.2\pm 0.6 $& 4 \\
           & Sy2 non-maser &$ 16     $&$ 1.1 \pm 0.9 $&$ 18     $&$ 1.3 \pm 0.5 $&$ 20     $&$ -0.2\pm 0.5 $& 24\\
           & Sy2 unknown   &$ 0      $&$ -           $&$ 1      $&$ 1.4         $&$ 1      $&$ -1.3        $& 1 \\
\multicolumn{1}{c}{}& & & & & & & \\
\hline                                                                                                      
\multicolumn{1}{c}{}& & & & & & & \\
12$\mu$m   & Sy1 maser     &$ 2      $&$1.25 \pm 0.93$&$ 2      $&$ 1.6 \pm 0.1 $&$ 2      $&$ 0.4 \pm 0.5 $& 2 \\
           & Sy1 non-maser &$ 22     $&$ 2.0 \pm 0.6 $&$ 22     $&$ 1.8 \pm 0.5 $&$ 27     $&$ 0.1 \pm 0.4 $& 30\\
           & Sy1 unknown   &$ 4      $&$ 2.3 \pm 0.6 $&$ 5      $&$ 1.9 \pm 0.6 $&$ 6      $&$ 0.0 \pm 0.3 $& 9 \\
           & Sy2 maser     &$ 11     $&$ 0.5 \pm 0.7 $&$ 10     $&$ 1.2 \pm 0.6 $&$ 11     $&$ -0.3\pm 0.5 $& 13\\
           & Sy2 non-maser &$ 29     $&$ 0.9 \pm 1.0 $&$ 34     $&$ 1.4 \pm 0.5 $&$ 42     $&$ -0.4\pm 0.6 $& 51\\
           & Sy2 unknown   &$ 1      $&$ 1.1         $&$ 1      $&$ 1.4         $&$ 2      $&$ -0.7\pm 0.9 $& 2 \\
\multicolumn{1}{c}{}& & & & & & & \\
\hline                                                                                                      
\multicolumn{1}{c}{}& & & & & & & \\
IRAS       & Sy1 maser     &$ 0      $&$ -           $&$ 0      $&$ -           $&$ 0      $&$ -           $& 0 \\
           & Sy1 non-maser &$ 13     $&$ 2.2 \pm 0.4 $&$ 10     $&$ 1.7 \pm 0.3 $&$ 15     $&$ 0.0 \pm 0.4 $& 15\\
           & Sy1 unknown   &$ 0      $&$ -           $&$ 0      $&$ -           $&$ 0      $&$ -           $& 0 \\
           & Sy2 maser     &$ 8      $&$ 0.2 \pm 1.0 $&$ 8      $&$ 1.2 \pm 0.7 $&$ 8      $&$ -0.3\pm 0.4 $& 8 \\
           & Sy2 non-maser &$ 6      $&$ 0.7 \pm 0.8 $&$ 9      $&$ 1.3 \pm 0.5 $&$ 10     $&$ -0.2\pm 0.5 $& 11\\
           & Sy2 unknown   &$ 0      $&$ -           $&$ 0      $&$ -           $&$ 0      $&$ -           $& 0 \\
\multicolumn{1}{c}{}& & & & & & & \\
\hline                                                                                                      
\multicolumn{1}{c}{}& & & & & & & \\
Combined:          & Sy1 maser     &$ 3      $&$ .8 \pm 0.9  $&$ 3      $&$ 1.6 \pm 0.1 $&$ 3      $&$ 0.0 \pm 0.6 $& 3 \\
CfA $\bigcup$      & Sy1 non-maser &$ 28     $&$ 2.0 \pm 0.6 $&$ 25     $&$ 1.8 \pm 0.5 $&$ 33     $&$ 0.1 \pm 0.4 $& 36\\
12$\mu$m $\bigcup$ & Sy1 unknown   &$ 5      $&$ 2.1 \pm 0.6 $&$ 6      $&$ 1.9 \pm 0.5 $&$ 7      $&$  0.0\pm 0.3 $& 10\\
IRAS  ~~           & Sy2 maser     &$ 13     $&$ 0.4 \pm 0.8 $&$ 12     $&$ 1.1 \pm 0.6 $&$ 13     $&$ -0.4\pm 0.5 $& 15\\
                   & Sy2 non-maser &$ 32     $&$ 0.9 \pm 1.0 $&$ 42     $&$ 1.4 \pm 0.5 $&$ 50     $&$ -0.3\pm 0.6 $& 60\\
                   & Sy2 unknown   &$ 1      $&$ 1.1         $&$ 1      $&$ 1.4         $&$ 2      $&$ -0.7\pm 0.9 $& 2 \\
\multicolumn{2}{c}{}  & \multicolumn{2}{c}{} & \multicolumn{2}{c}{} & \multicolumn{2}{c}{} \\
\hline
\multicolumn{2}{c}{}  & \multicolumn{2}{c}{} & \multicolumn{2}{c}{} & \multicolumn{2}{c}{} \\
  \end{tabular}
  \caption{{ The average values and standard deviations of the logarithmic luminosity ratios 
      for each subset of Seyfert galaxies. Column 1: In descending order, the optically selected 
      CfA sources \citep{1992ApJ...393...90H}, the MIR selected 12$\mu$m sources \citep{1993BAAS...25Q1362R}, 
      the IRAS F$25$/F$60$ flux-ratio selected sources \citep{2003ApJS..148..327S} and the combined sample
      that is used in this work. Columns 2-4: Number of sources with luminosities available in [O\,IV] and one of
      the following: 2-10 keV X-rays (Col. 2) or 7$\mu$m (Col. 3) or [O\,III] (Col. 4). Each sample's row 
      is subdivided into Sy1 maser, non-maser, maser-unknown and Sy2 maser, non-maser, maser-unknown.
      Column 5: Total Number of objects of the parent sample from Col. 1, e.g. 3 Sy1 maser in the CfA, 
      18 Sy1 non-maser in the CfA and so forth.
    }
  }
  \label{samplenumber_table}
  \end{center}
\end{table*}


\subsection{Comparison of the three samples with other known masers}\label{sample_properties}

{ 
Our combined sample was compiled from three complete 
samples with good coverage in the Spitzer archive.
Table \ref{complementation_table} shows the overlaps between the samples. 
Note that each sample is incomplete due to the limited availability of data  
in the Spitzer IRS archive and of X-Ray and [O\,III] measurements in the literature (see Tab. \ref{samplenumber_table}). Moreover, maser surveys 
were not performed with homogeneous properties (sensitivity, velocity coverage) and were not 
carried out for all sources of our combined sample.  
 
The three samples were based on different selection criteria: Optically selected Seyferts in the CfA sample 
and IRAS selected sources in the 12$\mu$m and F$25$/F$60$ sample. 
Thus, it is possible that they suffer from different 
biases with respect to potential maser detection.
The fraction of Sy2 masers to non-masers increases from $1/4$ (4/16) in the CfA sample, to $\sim 1/3$ (11/29)
in the 12$\mu$m sample and to $ \gtrsim 1$ (8/6) in the IRAS F25/F60 sample. 
This is consistent with the well known fact that the mid- and far-infrared wavelengths select more obscured AGN
than the optical bands.

However, the range of luminosity ratios (L$_{\rm{X}}$~/~L$_{\rm{[O\,IV]}}$, 
L$_{7\mu\rm{m}}$~/~L$_{\rm{[O\,IV]}}$, L$_{\rm{[O\,III]}}$~/~L$_{\rm{[O\,IV]}}$), listed in Tab. \ref{samplenumber_table}, 
are similar for all three samples.
This indicates that also among optical selected masing sources, some can be obscured with a level, 
similar to that of infrared selected sources\footnote{
Even for infrared selected AGN, optical criteria influence the sample, because the Seyfert identification 
is done by optical spectroscopy.
}.
To summarize, the result of all three samples (CfA, 12$\mu$m and IRAS F25/F60) are similar in that they point 
consistently to a prevalence of maser detections in Sy2s with high X-ray obscuration and 
one may expect this holds also for Seyfert galaxies in general.

Are our selected Sy2 masers representative for all 52 known 
Sy2 masers \citep{2010ApJ...708.1528Z}? 
To address this question, we 
compare our in-sample Sy2 masers with all remaining 37 off-sample Sy2 masers.

In Figure 
\ref{o4-Xo-msr_LUM_remove_off} we show a comparison  
of X-ray and [O\,IV] luminosities 
between in- and off-sample masers. 
The comparison refers to those masers with X-ray and [O\,IV] fluxes available, i.e. 12 off-sample Sy2s 13
in-sample Sy2s and 3 in-sample Sy1s.
Both, in-sample and off-sample roughly populate the same L$_{\rm{X}}$~/~L$_{\rm{[O\,IV]}}$ range. But the 
L$_{\rm{X}}$~/~L$_{\rm{[O\,IV]}}$ ratio is, on average, about a factor 2 higher for the off-sample than for the 
in-sample masers. This indicates that the off-sample Sy2 masers may be less absorbed than the in-sample ones.
Compared with the in-sample Sy2 non-masers (omitted in Fig \ref{o4-Xo-msr_LUM_remove_off}, 
see Fig. \ref{fig_o4_xo}), however, the off-sample masers show, on average, still about a 
factor 2 lower L$_{\rm{X}}$~/~L$_{\rm{[O\,IV]}}$, hence considerably high obscuration. 

Some off-sample masers show spurious flux ratios that imply no 
obscuration (i.e. L$_{\rm{X}}$~/~L$_{\rm{[O\,IV]}}~>~10$ for 4 objects). Among them, we find 
two nearby extended sources, NGC4258 and NGC4945, in which the X-ray emission has been associated 
with star formation by \citet{2002A&A...386..379R} and \citet{2004ApJS..151..193S}.

To summarize, the large overlap and the lack of significant differences between in- and off-sample Sy2 masers indicates 
that the results for our combined sample's Sy2 masers 
can be extended to all known Sy2 masers. 

We note that the inhomogeneous selection of all off-sample masers and non-masers precludes to derive a 
meaningful comparison of maser to non-maser statistics with our in-sample data.
A KS test shows a probability of 
63\% 
that both subsets, off- and in-sample masers, are drawn from the same parent distribution and the 
above mentioned difference in L$_{\rm{X}}$~/~L$_{\rm{[O\,IV]}}$ is only by chance. 
But similarly a KS test between the in-sample non-masers and the off-sample masers yields also 
60\% probability that they are drawn from the same parent distribution.
But yet, comparing the L$_{\rm{X}}$~/~L$_{\rm{[O\,IV]}}$ ratio between in-sample Sy1 non-masers and the
off-sample Sy2 masers, shows a probability of {0.16}\,\textperthousand~to be drawn from the same parent distribution.
This shows that the off-sample Sy2 masers are still significantly different from the unobscured Sy1 non-maser.

\begin{figure}
	\includegraphics[angle=0,width=\columnwidth]{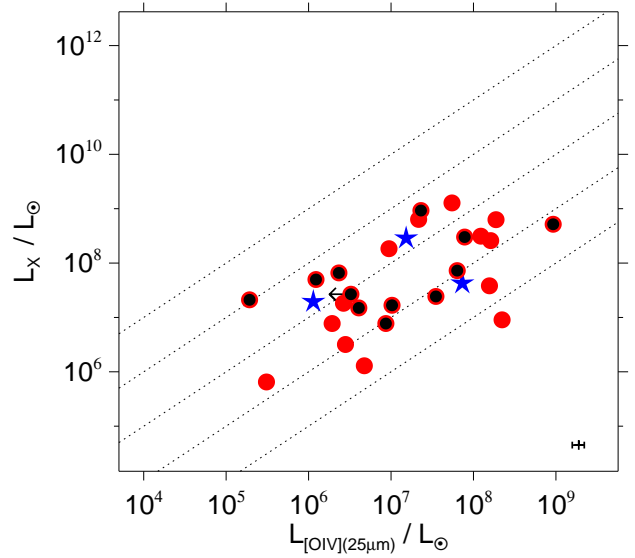}
	\caption{ {	Observed 2-10 keV X-rays plotted against [O\,IV] line luminosity.
            Sy1 masers are represented by blue stars and Sy2 masers by red dots. 
            Off-sample Sy2 masers are marked with black dots.
            The dotted lines mark fixed flux ratios of
	  1000; 100; 10; 1; 0.1 (from top to bottom). The error-bar in the lower 
          right corner is the average relative error of all [O\,IV]
          measurements.
	}
   	}
             
	\label{o4-Xo-msr_LUM_remove_off}
\end{figure}

\begin{figure}
	\includegraphics[angle=0,width=\columnwidth]{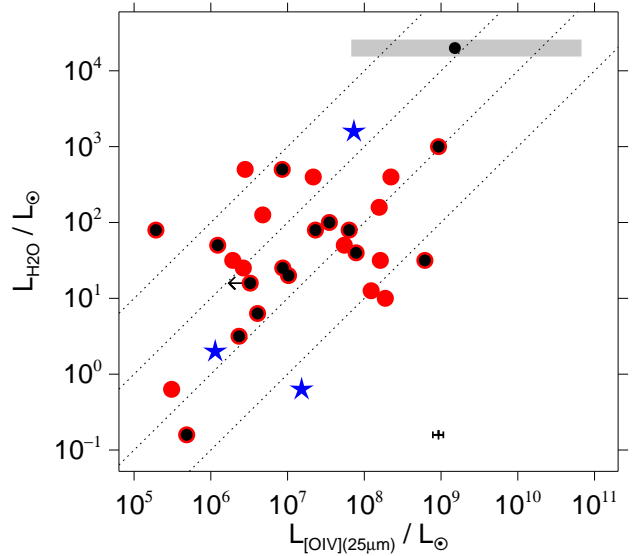}
	\caption{
          Maser H$_{\rm 2}$O versus [O\,IV] luminosity. 
                { Symbols and colors are as in 
                Fig. \ref{o4-Xo-msr_LUM_remove_off}.} 
                The dotted lines mark
                fixed-ratios 
                10$^{{-4}}$; 10$^{{-5}}$; 10$^{{-6}}$; 10$^{{-7}}$ 
                (from top to bottom).
                For comparison, the expected
                position of the $z = 0.66$  maser SDSSJ0804+3607 
                is marked with a black dot and a gray bar covering the
                range  
$~0.1 \,\times\, $L$_{\rm [O\,III]}~<~$L$_{\rm [O\,IV]}~<~100 \,\times\,
$L$_{\rm [O\,III]}$ assumed from Fig.~\ref{fig_o4_o3}. 
		}
	\label{o4-maser-msr_LUM}
\end{figure} 

\subsection{ Maser and AGN luminosity}
\label{section_luminosity}

A search for H$_{\rm 2}$O masers in 274 high-redshift ($0.3<z<0.8$) 
SDSS type-2 AGN half of which being type-2 \textit{quasars}   
(\citealt{2009ApJ...695..276B}) 
found only one maser 
(SDSSJ0804+3607, \citealt{2005ApJ...628L..89B}). 
The high rate of non-detections in these luminous AGN 
could be due to limited observational sensitivity or { to} intrinsic
differences between low- and high-luminosity AGN. 
Such differences could be, for instance, that in a high-luminosity AGN
the accretion disk becomes hotter so that the density required for
maser emission falls below a critical limit. 
If this is frequently the case, one would expect a { relative} decline of 
H$_{\rm 2}$O maser luminosity with increasing AGN luminosity. 
On the other hand, the SDSS H$_{\rm 2}$O maser survey was relatively
shallow, because one was interested to find masers which are
sufficiently bright for spatially resolved follow-up VLBI
observations.  

Here, we consider how far the Seyfert sample can help to distinguish
between these two possibilities { (i.e. by looking
whether or not L$_{\rm H2O}$~/~L$_{\rm [O\,IV]}$ declines with increasing L$_{\rm [O\,IV]}$)}.
A remarkable feature of Fig.~\ref{fig_o4_xo} is that 
maser-detections and non-detections are quite evenly
distributed along the whole [O\,IV] luminosity range covering about 
4 orders of magnitude. Thus our data do not indicate a 
trend that the frequency of non-masers rises with luminosity. 
We have also seen that the available 
maser observations of the Seyferts 
are biased against maser detection in faint (and distant) AGN 
(Fig.~\ref{Xo-o4_LUM_fx-sep_msr-HIST}).

Figure \ref{o4-maser-msr_LUM} shows the (isotropic) 
maser luminosity versus the AGN luminosity as traced by [O\,IV]. 
In addition to the Seyfert sample we have plotted the expected position
of SDSSJ0804+3607 at $z=0.66$, the only QSO-2 maser detection.
Because this source was not observed with Spitzer IRS, we
derived L$_{\rm [O\,IV]}$ from L$_{\rm [O\,III]}$ using the range 
0.1\,$\,\times\,$\,L$_{\rm [O\,III]}\,<\,$L$_{\rm [O\,IV]}\,<100 ~ \times\,$
L$_{\rm [O\,III]}$ as indicated in Fig.~\ref{fig_o4_o3}, which is also
valid for higher-luminosity AGN (\citealt{2005A&A...442L..39H}).
At a given [O\,IV] luminosity the maser luminosity spreads over three
orders of magnitude (Fig.~\ref{o4-maser-msr_LUM}). 
One explanation for the large spread is that the maser emission is, in
fact, not isotropic and hence the derived maser luminosity depends
sensitively on the maser direction with respect to the line-of-sight.
The Seyfert sample alone indicates only a marginal correlation
in Fig.~\ref{o4-maser-msr_LUM}, { with a Pearson correlation coefficient 
of $0.32$ for all Sy2 masers of our combined sample that is not significant 
at the $5\%$ level. Adding the off-sample Sy2 masers changes the coefficient to
$0.46$ which would then be significant,
{but the correlation could also be an artifact of distance in the luminosities.
However}, combined with 
the position of} SDSSJ0804+3607 and the fact, that its {assumed} H$_{\rm 2}$O /
[O\,IV] ratio lies in
the same range as for the lower luminosity AGN, argues in favor of a 
physical 
connection between maser and AGN luminosity. %

{ 
The numerous non-masers among the SDSS QSO-2s \citep{2009ApJ...695..276B} have a [O\,III] luminosity similar to that of SDSSJ0804+3607,
hence are expected to populate a similar L$_{\rm [O\,IV]}$ range in Fig. \ref{o4-maser-msr_LUM}.
The H$_{\rm 2}$O maser upper limits\footnote{ $\frac{{\rm L}_{ {\rm H2O}}}{{\rm L}_{\sun}}= 0.0039 \times \frac{1}{1+z} \times \left( \frac{{\rm D}_{\rm L}}{\rm Mpc}\right)^2$}, found for these QSO-2s by \citet{2009ApJ...695..276B}, 
lie even above J0804+3607.
}

Thus the upper limits are not stringent enough to support a relative 
decline of maser luminosity with increasing AGN luminosity. 
This, { together with the sufficiently high L$_{ {\rm H2O}}$~/~L$_{\rm [O\,IV]}$ ratio
of J0804+3607,} leads us to conclude that the main reason for the high rate of
maser non-detections is insufficient observational sensitivity, rather
than basic differences between low-and high-luminosity AGN for hosting a maser.


\section{Conclusion}
In order to understand the connection between H$_{\rm 2}$O 
maser detection rate and
nuclear extinction we used the [O\,IV]$_{25.9~\mu m}$ line and
the $7\,\mu$m continuum flux from Spitzer spectra of a well-selected
sample of 
{114} Seyfert galaxies, { from the CfA, 12$\mu$m and IRAS F25/F60 catalogs,} 
for which a maser search has been performed.   
These data were then compared to hard X-ray and [O\,III] 5007\AA~
fluxes from the literature. 
We analyzed the data in the framework of the orientation-dependent 
AGN unified scheme, yielding the following results: 
\begin{enumerate}
 \item 
  Comparing hard X-rays to [O\,IV] flux, Sy2s exhibit, on average, 
  an about 10 times lower X-ray  to [O\,IV] ratio than Sy1s. 
{ Masers prefer X-ray absorbed sources (i. e. low L$_{ {\rm X}}$~/~L$_{\rm [O\,IV]}$ ratios).  
 Sy2} masers present on average about { 4} 
times less X-ray flux normalized
  by [O\,IV] than non-maser { Sy2s}. 
  This is consistent with geometric alignment 
  of both the X-ray absorber and the 
  the maser emitting region in the accretion disk. 
  Non-masers do not show a preference for strongly absorbed
  sources. 
  However, our data indicate an observational bias 
  against faint sources, in the sense that 
  more sensitive { maser} observations might reveal more absorbed sources to
  house a maser. 
 \item 
  Regarding the 7\,$\mu$m to [O\,IV] flux ratio we find that most Sy2s
  spread along the same range as Sy1s. However there are sources with a
  significantly lower ratio, rendering the Sy2s on average about 3 times
  lower than Sy1s. These cases can be explained by an extended dusty
  absorber that is covering the 7\,$\mu$m emitting torus region. 
  Maser-detections also show a preference for 7\,$\mu$m  
  absorbed sources, but with less significance 
  than in the X-ray to
  [O\,IV] comparison. 
This suggests that the geometric alignment of the MIR
  absorber with the maser emitting disk is not as perfect 
  as the
  supposed alignment of the disk with the X-ray absorber. 
 \item 
  The [O\,III] to [O\,IV] flux ratio also indicates the presence of
  extended obscuration in some Sy2s that blocks the optical emission
  from the NLR. 
  Masers and non-masers are distributed very similarly in the [O\,III]
  to  [O\,IV] plot. The fraction of maser-detections is not
  significantly higher for sources with such extended absorption.  
  This leads us to conclude that the matter distribution for the
  [O\,III] absorber is not essential for the prediction of a maser detection. 
  Moreover, it
  is possible that a substantial fraction of the absorption of AGN
  emission could occur in extended regions outside the torus that are
  not necessarily aligned with the torus or AGN sub-structure. 
\item
  {
    The three samples, CfA, 12$\mu$m and IRAS F25/F60 
    provide very similar results.
    The Sy2 maser to non-maser fraction 
    increases from optical to infrared selection.
    The Sy2 masers of our combined sample 
    have a similar range of L$_{\rm [O\,IV]}$ and L$_{\rm H2O}$ as 
    the known off-sample
    Sy2 masers. While, on average, the off-sample masers are a factor two
    less obscured, 
    as inferred by
    the L$_{ {\rm X}}$~/~L$_{\rm [O\,IV]}$ ratio, they are still considerably obscured
    compared with Sy1s from our combined sample.
    Thus, the results obtained for
    our combined sample's Sy2 masers 
    may also hold for all remaining Sy2 maser
    sources 
    that have no Spitzer or X-ray data available.
}
 \item 
  After supplementing { 
  our combined sample with the remaining known Sy2 masers that were not
included in it}
  , 
  the H$_{\rm 2}$O maser luminosity appears to be correlated with 
  the AGN luminosity as traced by [O\,IV], although with a large
  spread. We do not find evidence for physical differences between
  low-and high-luminosity AGN for housing a maser.

\end{enumerate}
The results demonstrate that heavy X-ray absorption is an indicator
for high probability to detect a maser. 
The $7\,\mu$m absorption can also be used to find maser candidates,
but with 
lower probability.

\begin{acknowledgements}
  The work is based on observations made with the {\it Spitzer
    Space Telescope,}\/ which is operated by the Jet Propulsion Laboratory,
  California Institute of Technology under a contract with NASA.
  This research has made use of the NASA/IPAC Extragalactic Database
  (NED) which is operated by the Jet Propulsion Laboratory, California
  Institute of Technology, under contract with the National
  Aeronautics and Space Administration. 
  This publication is supported as a project of the  Nordrhein-Westf\"alische Akademie der Wissenschaften und der K\"unste in the framework of the  academy program by the Federal Republic of Germany and the state Nordrhein-Westfalen.
  We thank the referee J. S. Zhang for his careful review
  of the manuscript.

\end{acknowledgements}


%
\bibliographystyle{aa} 
\bibliography{maser_aanda}

\twocolumn
\scriptsize{
\tablefirsthead{%
\hline
& & &   & & & \\
~~~(1)& (2) & (3) & (4) & (5)& (6) & (7) \\
& & &   & & & \\
Source &{ H$_2$O} & Sy&  [O\,IV] &[O\,III]& { 2-10~keV} &  $7.6~\mu$m  \\ 
\cmidrule(r){4-5}\cmidrule(r){6-7}
& log $\left( \frac{\rm{L}}{\rm{L}_{\sun}} \right)$ & & \multicolumn{2}{c}{$\left( 10^{-15}~\frac{\mbox{erg}}{\mbox{s cm}^{2}}\right)$}&\multicolumn{2}{c}{$\left( 10^{-14}~\frac{\mbox{erg}}{\mbox{s cm}^{2}}\right)$} \\
& & &   & & & \\
\hline
& & &   & & & \\
}
\tablehead{%
\hline
~~~~~~~~~&~~~~~~~ &~~~ & ~~~~~~~~~  &~~~~~~~~~ & ~~~~~~~~~&~~~~~~~~~ \\
  Source & { H$_2$O} & Sy&  [O\,IV] &[O\,III]&{ 2-10~keV}&  $7.6~\mu$m  \\ 
& & &   & & & \\
\hline
& & &   & & & \\
}
\tablelasttail{%
& & &   & & & \\
\hline
& & &   & & & \\
}
\bottomcaption[]{Measured fluxes and literature values. 
   Column 1: { Source names with catalog marks: \\
\textit{a}: CfA sample \\
\textit{b}: 12$\mu$m sample \\
\textit{c}: IRAS F25/F60 sample\\ }
  Column 2: { Isotropic maser luminosities obtained from \citealt{2009ApJ...695..276B}. 
    A question mark designates sources unobserved for masers and a dash represents a maser undetected source.  \\
  Note that MRK938 and NGC1320 are listed as Maser in \citet{2010ApJ...708.1528Z} 
but have no luminosity information available. Therefore, they are marked with asterisks. }
  Column 3: Seyfert type obtained from the NED or the literature with references given in square brackets.
  Column 4: [O\,IV] flux determined by
  \citet{Ramolla:Thesis:2009}. Column 5: [O\,III] flux from the literature.%
  Column 6: 2-10 keV X-ray flux obtained
  from the literature. 
  Column 7: $7\,\mu$m continuum flux determined by
  \citet{Ramolla:Thesis:2009}. 
  { An analysis of the off-sample masers has also been performed. This data is appended in the table.\\
References:\\       1: \citealt{1988ApJS...67..249D}\\       2: \citealt{1996ApJS..106..399P}\\       3: \citealt{1992ApJS...79...49W}\\       4: \citealt{2002A&A...388...74G}\\       5: \citealt{2006ApJS..164...81M}\\       6: \citealt{2007ApJ...657..167S}\\       7: \citealt{2003ApJ...583..632T}\\       8: \citealt{1992A&AS...96..389D}\\       9: \citealt{2000ApJ...542..175A}\\      10: \citealt{2000MNRAS.316..234R}\\      11: \citealt{2005ApJS..161..185U}\\      12: \citealt{2003NewAR..47.1091T}\\      13: \citealt{2005A&A...444..119G}\\      14: \citealt{2007ApJ...668...81M}\\      15: \citealt{1989MNRAS.240..833T}\\      16: \citealt{2006A&A...446..459C}\\      17: \citealt{1996PASJ...48..231I}\\      18: \citealt{2006yCat..73660480G}\\      19: \citealt{2003ApJS..148..327S}\\      20: \citealt{1997AJ....114.1345V}\\      21: \citealt{2008ApJ...686L..13G}\\      22: \citealt{2006A&A...455..773V}\\      23: \citealt{2000A&A...357...13R}\\      24: \citealt{2006A&A...446..919B}\\      25: \citealt{1992ApJ...397..442B}\\      26: \citealt{2007A&A...472..705V}\\      27: \citealt{1992MNRAS.255..197R}\\      28: \citealt{1995ApJS...98..103S}\\      29: \citealt{1999ApJS..121..473B}\\      30: \citealt{1995ApJS...98..129K}\\      31: \citealt{2001A&A...368...44S}\\      32: \citealt{2005MNRAS.360..380B}\\      33: \citealt{2003AJ....126..153I}\\      34: \citealt{1995ApJS...98..477H}\\      35: \citealt{2002ApJS..139....1T}\\      36: \citealt{2006A&A...460...45G}\\      37: \citealt{2007MNRAS.382..194N}\\      38: \citealt{2003A&A...397..883G}\\      39: \citealt{2002A&A...392..453B}\\      40: \citealt{2000A&A...363..863M}\\      41: \citealt{2007AJ....134..294S}\\      42: \citealt{2002A&A...389..802P}\\      43: \citealt{2007MNRAS.375..227M}\\      44: \citealt{1983ApJ...266..485P}\\      45: \citealt{2007ApJ...663..799H}\\      46: \citealt{2001A&A...378..806G}\\      47: \citealt{2006ApJ...648..111L}\\      48: \citealt{2003A&A...402..141B}\\      49: \citealt{1997MNRAS.286..513R}\\      50: \citealt{2003A&A...398..967G}\\      51: \citealt{2006AJ....132..401B}\\      52: \citealt{2003A&A...398..107B}\\      53: \citealt{2000A&A...358..117P}\\      54: \citealt{2001ApJS..132...37K}\\      55: \citealt{1997MNRAS.288..920L}\\      56: \citealt{1984MNRAS.208...15S}\\      57: \citealt{2000AdSpR..25..823U}\\      58: \citealt{2006AJ....131.2843S}\\      59: \citealt{1997ApJS..112..315H}\\      60: \citealt{2004AJ....127..606W}\\      61: \citealt{2009SciChin.G6.960}\\      62: \citealt{2008ApJ...678..701T}\\      63: \citealt{1980A&AS...40..295H}\\      64: \citealt{2003ApJ...588..763D}\\      65: \citealt{1994A&A...288..457O}\\      66: \citealt{2001ApJ...557..180S}\\       67: \citealt{2001MNRAS.328L..32L}\\      68: \citealt{2004MNRAS.348.1451L}\\      69: \citealt{2004ApJ...617..930M}\\      70: \citealt{1997ApJS..113...23T}\\      71: \citealt{2006A&A...455..173P}\\
  }
 }
  \label{table1}
\begin{supertabular}{@{}l@{}c@{\hspace{2mm}}c@{\hspace{2mm}}c@{\hspace{2mm}}c@{\hspace{2mm}}c@{\hspace{2mm}}c@{}}
%
              MRK334$~^{[a]}$&$  - $&$ 1.8 $&$                     82 \pm    14 $&$          49~^{\bf [1]}          $&$         800~^{\bf [2]}          $&$                    504 \pm   101  $\\
            MRK335$~^{[a,b]}$&$  - $&$ 1.0 $&$                     67 \pm    10 $&$         950~^{\bf [3]}          $&$         960~^{\bf [4]}          $&$                    983 \pm   241  $\\
              MRK938$~^{[b]}$&$ * $&$ 2.0 $&$                               -  $&$          44~^{\bf [5]}          $&$          23~^{\bf [6]}          $&$                    332 \pm    66  $\\
             E12-G21$~^{[b]}$&$  ? $&$ 1.0 $&$                    187 \pm    56 $&$          97~^{\bf [7]}          $&$                               -  $&$                    366 \pm    82  $\\
            MRK348$~^{[b,c]}$&$ 2.6$&$ 2.0 $&$                    163 \pm    25 $&$         359~^{\bf [8]}          $&$         482~^{\bf [9]}          $&$                   1151 \pm   235  $\\
                IZw1$~^{[b]}$&$  ? $&$ 1.0 $&$                     97 \pm    13 $&$          44~^{\bf [1]}          $&$         680~^{\bf [10]}         $&$                        -          $\\
      IRAS00521-7054$~^{[b]}$&$  - $&$ 2.0 $&$                     71 \pm    11 $&$          77~^{\bf [8]}          $&$                               -  $&$                        -          $\\
              NGC424$~^{[b]}$&$  - $&$ 2.0 $&$                    223 \pm    34 $&$         420~^{\bf [3]}          $&$         122~^{\bf [11]}         $&$                   3655 \pm   738  $\\
             NGC526A$~^{[b]}$&$  - $&$ 1.5 $&$                    176 \pm    26 $&$         270~^{\bf [3]}          $&$        2046~^{\bf [11]}         $&$                    669 \pm   162  $\\
              NGC513$~^{[b]}$&$  - $&$ 2.0 $&$                     59 \pm     9 $&$          35~^{\bf [12]}         $&$                               -  $&$                    232 \pm    59  $\\
       F01475-0740$~^{[b,c]}$&$  - $&$ 2.0 $&$                     62 \pm    10 $&$          53~^{\bf [8]}          $&$          82~^{\bf [13]}         $&$                    393 \pm   111  $\\
               UM146$~^{[a]}$&$  - $&$ 1.9 $&$                     26 \pm     3 $&$          60~^{\bf [3]}          $&$                               -  $&$                          <   199  $\\
              MRK590$~^{[c]}$&$  - $&$ 1.2 $&$                     31 \pm     8 $&$          53~^{\bf [1]}          $&$        1970~^{\bf [14]}         $&$                          <   589  $\\
     MCG+05-06-036$~^{[a,b]}$&$  ? $&$ 1.0 $&$                     42 \pm     5 $&$                            -     $&$                               -  $&$                    166 \pm    33  $\\
            NGC931$~^{[b,c]}$&$  - $&$ 1.5 $&$                    459 \pm    67 $&$          75~^{\bf [8]}          $&$        2000~^{\bf [15]}         $&$                   1697 \pm   392  $\\
         NGC1068$~^{[a,b,c]}$&$ 2.2$&$ 2.0 $&$                  18908 \pm  2697 $&$        4834~^{\bf [8]}          $&$         462~^{\bf [16]}         $&$                  52585 \pm 10567  $\\
             NGC1056$~^{[b]}$&$  - $&$ 2.0 $&$                          <   212 $&$          23~^{\bf [7]}          $&$                               -  $&$                    235 \pm    47  $\\
           NGC1097$~^{[a,b]}$&$  - $&$ 1.0 $&$                     52 \pm    12 $&$          18~^{\bf [7]}          $&$         170~^{\bf [17]}         $&$                    283 \pm    58  $\\
             NGC1125$~^{[b]}$&$  - $&$ 2.0 $&$                    356 \pm    52 $&$          23~^{\bf [18]}         $&$                               -  $&$                    118 \pm    27  $\\
             NGC1144$~^{[b]}$&$  - $&$ 2.0 $&$                     69 \pm    10 $&$                            -     $&$       11000~^{\bf [2]}          $&$                    164 \pm    37  $\\
          M-2-8-39$~^{[b,c]}$&$  - $&$ 2.0 $&$                    144 \pm    21 $&$         183~^{\bf [8]}          $&$                               -  $&$                    270 \pm    69  $\\
           NGC1194$~^{[b,c]}$&$  - $&$ 1.0 $&$                    144 \pm    21 $&$         396~^{\bf [19]}         $&$                               -  $&$                    521 \pm   105  $\\
             NGC1241$~^{[b]}$&$  - $&$ 2.0 $&$                          <   100 $&$         370~^{\bf [20]}         $&$                               -  $&$                        -          $\\
           NGC1320$~^{[b,c]}$&$ * $&$ 2.0 $&$                    254 \pm    37 $&$         122~^{\bf [8]}          $&$         496~^{\bf [21]}         $&$                    933 \pm   231  $\\
             NGC1365$~^{[b]}$&$  - $&$ 1.8 $&$                   1441 \pm   207 $&$          62~^{\bf [22]}         $&$         660~^{\bf [23]}         $&$                   2759 \pm   553  $\\
           NGC1386$~^{[b,c]}$&$ 2.1$&$ 2.0 $&$                    991 \pm   145 $&$         800~^{\bf [24]}         $&$          27~^{\bf [18]}         $&$                   1017 \pm   206  $\\
      IRAS03362-1641$~^{[b]}$&$  - $&$ 2.0 $&$                     52 \pm     8 $&$          18~^{\bf [8]}          $&$                               -  $&$                        -          $\\
         F03450+0055$~^{[b]}$&$  ? $&$ 1.5 $&$                     31 \pm     5 $&$         100~^{\bf [25]}         $&$                               -  $&$                          < 10504  $\\
             3C120$~^{[a,b]}$&$  - $&$ 1.0 $&$                   1195 \pm   174 $&$         304~^{\bf [8]}          $&$        8200~^{\bf [26]}         $&$                    987 \pm   235  $\\
              MRK618$~^{[b]}$&$  - $&$ 1.0 $&$                     96 \pm    16 $&$         160~^{\bf [8]}          $&$         700~^{\bf [27]}         $&$                        -          $\\
         F04385-0828$~^{[b]}$&$  - $&$ 2.0 $&$                     80 \pm    14 $&$           3~^{\bf [7]}          $&$        1800~^{\bf [2]}          $&$                   1119 \pm   228  $\\
             NGC1667$~^{[b]}$&$  - $&$ 2.0 $&$                     68 \pm    11 $&$          64~^{\bf [28]}         $&$           3~^{\bf [29]}         $&$                     76 \pm    18  $\\
            E33-G2$~^{[b,c]}$&$  - $&$ 2.0 $&$                    137 \pm    20 $&$          57~^{\bf [19]}         $&$                               -  $&$                        -          $\\
         M-5-13-17$~^{[b,c]}$&$  - $&$ 1.5 $&$                     98 \pm    15 $&$         340~^{\bf [19]}         $&$                               -  $&$                    376 \pm    96  $\\
      IRAS05189-2524$~^{[b]}$&$  - $&$ 2.0 $&$                    218 \pm    16 $&$          39~^{\bf [30]}         $&$         360~^{\bf [31]}         $&$                   2247 \pm   451  $\\
          Markarian3$~^{[c]}$&$ 1.0$&$ 2.0 $&$                   1763 \pm   358 $&$        1070~^{\bf [8]}          $&$         590~^{\bf [32]}         $&$                   1593 \pm   349  $\\
              MRK6$~^{[b,c]}$&$  - $&$ 1.5 $&$                    385 \pm    56 $&$         700~^{\bf [8]}          $&$        1200~^{\bf [33]}         $&$                        -          $\\
                MRK9$~^{[b]}$&$  - $&$ 1.5 $&$                     48 \pm     8 $&$         109~^{\bf [3]}          $&$                               -  $&$                          <  1944  $\\
             MRK79$~^{[b,c]}$&$  - $&$ 1.2 $&$                    395 \pm    57 $&$         370~^{\bf [3]}          $&$        2600~^{\bf [15]}         $&$                          <  3567  $\\
    IRAS07598+6508$~^{[a,b]}$&$  ? $&$ 1.0 $&$                          <   168 $&$                            -     $&$                               -  $&$                        -          $\\
              MRK622$~^{[c]}$&$  - $&$ 2.0 $&$                     66 \pm     8 $&$          40~^{\bf [19]}         $&$          22~^{\bf [13]}         $&$                        -          $\\
             NGC2639$~^{[b]}$&$ 1.4$&$ 1.9 $&$                     36 \pm     4 $&$          14~^{\bf [34]}         $&$          25~^{\bf [35]}         $&$                          <   155  $\\
    IRAS08572+3915$~^{[a,b]}$&$  ? $&$ 2.0 $&$                    167 \pm    50 $&$           8~^{\bf [5]}          $&$                               -  $&$                    427 \pm    85  $\\
              MRK704$~^{[b]}$&$  - $&$ 1.5 $&$                    117 \pm    18 $&$          85~^{\bf [1]}          $&$         537~^{\bf [11]}         $&$                          < 10595  $\\
             NGC2841$~^{[a]}$&$  - $&$ 1.0 $&$                     12 \pm     3 $&$                            -     $&$                               -  $&$                    161 \pm    51  $\\
          pg0923+129$~^{[c]}$&$  - $&$ 1.2 $&$                     74 \pm    12 $&$          90~^{\bf [19]}         $&$        1151~^{\bf [11]}         $&$                    458 \pm    96  $\\
             UGC5101$~^{[a]}$&$ 3.2$&$ 1.5 $&$                     82 \pm    11 $&$          21~^{\bf [5]}          $&$           5~^{\bf [36]}         $&$                    276 \pm    55  $\\
           NGC2992$~^{[a,b]}$&$  - $&$ 1.9 $&$                   1300 \pm   134 $&$         360~^{\bf [1]}          $&$        8030~^{\bf [37]}         $&$                    639 \pm   130  $\\
             MRK1239$~^{[b]}$&$  - $&$ 1.5 $&$                    154 \pm    24 $&$         467~^{\bf [8]}          $&$                               -  $&$                   3323 \pm   672  $\\
           NGC3031$~^{[a,b]}$&$  - $&$ 1.8 $&$                     44 \pm    13 $&$         100~^{\bf [34]}         $&$        1500~^{\bf [29]}         $&$                        -          $\\
               3C234$~^{[b]}$&$  ? $&$ 1.0 $&$                     79 \pm    12 $&$                            -     $&$                               -  $&$                    407 \pm    92  $\\
           NGC3079$~^{[a,b]}$&$ 2.7$&$ 2.0 $&$                    290 \pm    53 $&$         945~^{\bf [5]}          $&$          33~^{\bf [16]}         $&$                    160 \pm    32  $\\
           NGC3227$~^{[a,b]}$&$  - $&$ 1.5 $&$                    655 \pm    95 $&$         820~^{\bf [1]}          $&$         750~^{\bf [38]}         $&$                        -          $\\
             NGC3281$~^{[c]}$&$  - $&$ 2.0 $&$                   1779 \pm   534 $&$          55~^{\bf [1]}          $&$                               -  $&$                    162 \pm    32  $\\
             NGC3393$~^{[c]}$&$ 2.6$&$ 2.0 $&$                   2214 \pm   184 $&$         268~^{\bf [18]}         $&$           9~^{\bf [18]}         $&$                    199 \pm    52  $\\
             NGC3511$~^{[b]}$&$  - $&$ 1.0 $&$                     23 \pm     6 $&$                            -     $&$                               -  $&$                     27 \pm     5  $\\
           NGC3516$~^{[a,c]}$&$  - $&$ 1.5 $&$                    451 \pm    66 $&$         270~^{\bf [1]}          $&$        4410~^{\bf [18]}         $&$                          <  2900  $\\
           M+0-29-23$~^{[b]}$&$  - $&$ 2.0 $&$                     78 \pm    23 $&$           5~^{\bf [7]}          $&$                               -  $&$                    348 \pm    69  $\\
             NGC3660$~^{[b]}$&$  - $&$ 2.0 $&$                     25 \pm     5 $&$          33~^{\bf [28]}         $&$                               -  $&$                          <   234  $\\
           NGC3783$~^{[a,c]}$&$  - $&$ 1.0 $&$                    378 \pm    57 $&$         763~^{\bf [8]}          $&$        8500~^{\bf [39]}         $&$                   2261 \pm   470  $\\
             NGC3786$~^{[a]}$&$  - $&$ 1.8 $&$                    129 \pm    19 $&$          84~^{\bf [3]}          $&$                               -  $&$                    281 \pm    56  $\\
           NGC3982$~^{[a,b]}$&$  - $&$ 2.0 $&$                     89 \pm    15 $&$         188~^{\bf [34]}         $&$          22~^{\bf [11]}         $&$                     49 \pm    10  $\\
           NGC4051$~^{[a,b]}$&$ 0.3$&$ 1.5 $&$                    366 \pm    53 $&$         390~^{\bf [34]}         $&$         627~^{\bf [16]}         $&$                   1704 \pm   344  $\\
           UGC7064$~^{[a,b]}$&$  - $&$ 1.9 $&$                    118 \pm    17 $&$                            -     $&$                               -  $&$                    269 \pm    55  $\\
           NGC4151$~^{[a,b]}$&$-0.2$&$ 1.5 $&$                   2396 \pm   342 $&$       11600~^{\bf [34]}         $&$        4510~^{\bf [16]}         $&$                   7211 \pm  1459  $\\
          MRK766$~^{[a,b,c]}$&$  - $&$ 1.0 $&$                    474 \pm    69 $&$         453~^{\bf [8]}          $&$        3000~^{\bf [40]}         $&$                   1061 \pm   213  $\\
         NGC4388$~^{[a,b,c]}$&$ 1.1$&$ 2.0 $&$                   2996 \pm   644 $&$         564~^{\bf [8]}          $&$         762~^{\bf [16]}         $&$                    971 \pm   199  $\\
               3C273$~^{[b]}$&$  ? $&$ 1.0 $&$                     79 \pm     9 $&$         116~^{\bf [41]}         $&$        8300~^{\bf [15]}         $&$                   2043 \pm   424  $\\
           NGC4501$~^{[a,b]}$&$  - $&$ 2.0 $&$                     33 \pm     6 $&$          34~^{\bf [34]}         $&$          11~^{\bf [16]}         $&$                          <   271  $\\
           NGC4507$~^{[a,c]}$&$  - $&$ 2.0 $&$                    332 \pm    51 $&$         828~^{\bf [8]}          $&$        2100~^{\bf [29]}         $&$                   1579 \pm   320  $\\
             NGC4569$~^{[a]}$&$  - $&$ 2.0 $&$                     42 \pm    10 $&$          24~^{\bf [5]}          $&$                               -  $&$                    463 \pm    95  $\\
           NGC4579$~^{[a,b]}$&$  - $&$ 1.9 $&$                     30 \pm     5 $&$                            -     $&$         440~^{\bf [29]}         $&$                    325 \pm    74  $\\
           NGC4593$~^{[b,c]}$&$  - $&$ 1.0 $&$                    127 \pm    40 $&$         134~^{\bf [8]}          $&$        3710~^{\bf [42]}         $&$                   1428 \pm   304  $\\
             NGC4602$~^{[b]}$&$  - $&$ 1.9 $&$                          <    66 $&$         134~^{\bf [8]}          $&$                               -  $&$                     70 \pm    18  $\\
         TOL1238-364$~^{[b]}$&$  - $&$ 2.0 $&$                    145 \pm    23 $&$         194~^{\bf [18]}         $&$          17~^{\bf [11]}         $&$                    779 \pm   161  $\\
           M-2-33-34$~^{[b]}$&$  - $&$ 1.0 $&$                    670 \pm   145 $&$         364~^{\bf [8]}          $&$                               -  $&$                    175 \pm    57  $\\
              MRK231$~^{[b]}$&$  - $&$ 1.0 $&$                    233 \pm    70 $&$         230~^{\bf [1]}          $&$          68~^{\bf [11]}         $&$                   5960 \pm  1192  $\\
             NGC4826$~^{[a]}$&$  - $&$ 2.0 $&$                    139 \pm    41 $&$                            -     $&$                               -  $&$                    540 \pm   108  $\\
             NGC4922$~^{[b]}$&$ 2.3$&$ 2.0 $&$                               -  $&$          64~^{\bf [5]}          $&$                               -  $&$                    476 \pm    96  $\\
             NGC4941$~^{[b]}$&$  - $&$ 2.0 $&$                    285 \pm    43 $&$         143~^{\bf [18]}         $&$          82~^{\bf [11]}         $&$                          <   513  $\\
           NGC4968$~^{[b,c]}$&$  - $&$ 2.0 $&$                    307 \pm    45 $&$         177~^{\bf [8]}          $&$          15~^{\bf [13]}         $&$                    742 \pm   156  $\\
           NGC5005$~^{[a,b]}$&$  - $&$ 2.0 $&$                    179 \pm    21 $&$           7~^{\bf [1]}          $&$                               -  $&$                    344 \pm    69  $\\
           NGC5033$~^{[a,b]}$&$  - $&$ 1.8 $&$                    109 \pm    23 $&$          53~^{\bf [1]}          $&$         550~^{\bf [29]}         $&$                    211 \pm    43  $\\
     MCG-03-34-064$~^{[a,b]}$&$  - $&$ 1.8 $&$                   1062 \pm   153 $&$        1507~^{\bf [8]}          $&$         210~^{\bf [43]}         $&$                   2206 \pm   442  $\\
             NGC5135$~^{[b]}$&$  - $&$ 2.0 $&$                    726 \pm   147 $&$         219~^{\bf [44]}         $&$                               -  $&$                    808 \pm   161  $\\
             NGC5194$~^{[b]}$&$-0.2$&$ 2.0 $&$                    227 \pm    47 $&$         120~^{\bf [34]}         $&$          48~^{\bf [16]}         $&$                    199 \pm    46  $\\
         M-6-30-15$~^{[b,c]}$&$  - $&$ 1.2 $&$                    227 \pm    34 $&$          75~^{\bf [19]}         $&$        4220~^{\bf [37]}         $&$                   1425 \pm   313  $\\
    IRAS13349+2438$~^{[a,b]}$&$  ? $&$ 1.0 $&$                     64 \pm     9 $&$          47~^{\bf [7]}          $&$         360~^{\bf [45]}         $&$                   3194 \pm   673  $\\
              MRK266$~^{[b]}$&$ 1.5$&$ 2.0 $&$                    349 \pm    77 $&$          23~^{\bf [34]}         $&$          56~^{\bf [23]}         $&$                     72 \pm    14  $\\
            MRK273$~^{[a,b]}$&$  - $&$ 2.0 $&$                    474 \pm   142 $&$         213~^{\bf [5]}          $&$          60~^{\bf [29]}         $&$                        -          $\\
           IC4329a$~^{[a,b]}$&$  - $&$ 1.2 $&$                   1061 \pm   156 $&$         340~^{\bf [12]}         $&$       16400~^{\bf [46]}         $&$                   4140 \pm   968  $\\
         NGC5347$~^{[a,b,c]}$&$ 1.5$&$ 2.0 $&$                     54 \pm     9 $&$          45~^{\bf [8]}          $&$          22~^{\bf [47]}         $&$                    628 \pm   129  $\\
           MRK463E$~^{[a,b]}$&$  - $&$ 2.0 $&$                    639 \pm    96 $&$         563~^{\bf [1]}          $&$          40~^{\bf [29]}         $&$                   2295 \pm   461  $\\
             NGC5506$~^{[b]}$&$ 1.7$&$ 1.9 $&$                   2492 \pm   360 $&$         521~^{\bf [28]}         $&$        5800~^{\bf [48]}         $&$                   4222 \pm   844  $\\
         NGC5548$~^{[a,b,c]}$&$  - $&$ 1.5 $&$                    141 \pm    24 $&$         360~^{\bf [8]}          $&$        4300~^{\bf [49]}         $&$                    726 \pm   174  $\\
            MRK817$~^{[a,b]}$&$  - $&$ 1.5 $&$                     73 \pm    12 $&$         140~^{\bf [1]}          $&$                               -  $&$                    950 \pm   244  $\\
          PG1501+106$~^{[a]}$&$  - $&$ 1.5 $&$                    246 \pm    36 $&$         250~^{\bf [1]}          $&$        1869~^{\bf [11]}         $&$                        -          $\\
           NGC5929$~^{[a,b]}$&$  - $&$ 2.0 $&$                          <   114 $&$          93~^{\bf [1]}          $&$         197~^{\bf [11]}         $&$                     32 \pm     8  $\\
             NGC5953$~^{[b]}$&$  - $&$ 2.0 $&$                    172 \pm    25 $&$          63~^{\bf [3]}          $&$                               -  $&$                    259 \pm    52  $\\
            M-2-40-4$~^{[b]}$&$  - $&$ 2.0 $&$                    115 \pm    19 $&$          74~^{\bf [12]}         $&$        2693~^{\bf [11]}         $&$                   1586 \pm   382  $\\
         F15480-0344$~^{[b]}$&$  - $&$ 2.0 $&$                    364 \pm    53 $&$         138~^{\bf [8]}          $&$          37~^{\bf [13]}         $&$                          <   838  $\\
         ESO141-G055$~^{[b]}$&$  ? $&$ 1.0 $&$                    107 \pm    16 $&$         164~^{\bf [8]}          $&$        2650~^{\bf [50]}         $&$                        -          $\\
    IRAS19254-7245$~^{[a,b]}$&$  - $&$ 2.0 $&$                    105 \pm    31 $&$         602~^{\bf [51]}         $&$          20~^{\bf [52]}         $&$                          <   323  $\\
             NGC6810$~^{[b]}$&$  - $&$ 2.0 $&$                     68 \pm    13 $&$          13~^{\bf [7]}          $&$                               -  $&$                    838 \pm   168  $\\
           NGC6860$~^{[b,c]}$&$  - $&$ 1.0 $&$                    122 \pm    18 $&$          25~^{\bf [19]}         $&$        4900~^{\bf [26]}         $&$                          <  2201  $\\
             NGC6890$~^{[b]}$&$  - $&$ 2.0 $&$                     90 \pm    13 $&$          72~^{\bf [18]}         $&$           8~^{\bf [11]}         $&$                    410 \pm    97  $\\
            MRK509$~^{[a,b]}$&$  - $&$ 1.2 $&$                    286 \pm    44 $&$         540~^{\bf [8]}          $&$        5660~^{\bf [53]}         $&$                   1221 \pm   254  $\\
            UGC11630$~^{[c]}$&$  - $&$ 2.0 $&$                    175 \pm    28 $&$                            -     $&$                               -  $&$                    280 \pm    77  $\\
          IC5063$~^{[a,b,c]}$&$  - $&$ 2.0 $&$                   1139 \pm   167 $&$         564~^{\bf [8]}          $&$        1200~^{\bf [29]}         $&$                   2949 \pm   598  $\\
            UGC11680$~^{[b]}$&$  - $&$ 2.0 $&$                     45 \pm    13 $&$          88~^{\bf [54]}         $&$                               -  $&$                          <   686  $\\
          PG2130+099$~^{[a]}$&$  ? $&$ 1.0 $&$                    103 \pm    16 $&$         104~^{\bf [8]}          $&$         530~^{\bf [55]}         $&$                    861 \pm   277  $\\
              IC5135$~^{[b]}$&$  - $&$ 2.0 $&$                    300 \pm    40 $&$          27~^{\bf [30]}         $&$           6~^{\bf [36]}         $&$                    451 \pm    90  $\\
             NGC7172$~^{[b]}$&$  - $&$ 2.0 $&$                    384 \pm    39 $&$          10~^{\bf [56]}         $&$        2200~^{\bf [6]}          $&$                    522 \pm   104  $\\
      IRAS22017+0319$~^{[b]}$&$  ? $&$ 2.0 $&$                    287 \pm    42 $&$         218~^{\bf [8]}          $&$         360~^{\bf [57]}         $&$                        -          $\\
         NGC7213$~^{[a,b,c]}$&$  - $&$ 1.5 $&$                     45 \pm     8 $&$         130~^{\bf [8]}          $&$        3660~^{\bf [14]}         $&$                    798 \pm   202  $\\
             3C445$~^{[a,b]}$&$  - $&$ 1.0 $&$                     71 \pm    14 $&$                            -     $&$         700~^{\bf [58]}         $&$                    765 \pm   242  $\\
           NGC7314$~^{[a,b]}$&$  - $&$ 1.9 $&$                    690 \pm   101 $&$          61~^{\bf [3]}          $&$        3560~^{\bf [29]}         $&$                    249 \pm    65  $\\
          UGC12138$~^{[a,c]}$&$  - $&$ 1.8 $&$                    105 \pm    15 $&$         144~^{\bf [8]}          $&$                               -  $&$                    273 \pm    57  $\\
            M-3-58-7$~^{[b]}$&$  - $&$ 2.0 $&$                    117 \pm    19 $&$         251~^{\bf [7]}          $&$                               -  $&$                   1186 \pm   272  $\\
           NGC7469$~^{[a,b]}$&$  - $&$ 1.2 $&$                    322 \pm    48 $&$         840~^{\bf [1]}          $&$        2900~^{\bf [49]}         $&$                   2298 \pm   460  $\\
           NGC7582$~^{[a,b]}$&$  - $&$ 2.0 $&$                   2449 \pm   587 $&$         300~^{\bf [1]}          $&$        1550~^{\bf [29]}         $&$                    309 \pm    61  $\\
             NGC7590$~^{[b]}$&$  - $&$ 2.0 $&$                     58 \pm    18 $&$          11~^{\bf [18]}         $&$                               -  $&$                     70 \pm    17  $\\
           NGC7603$~^{[a,b]}$&$  - $&$ 1.5 $&$                     24 \pm     4 $&$          29~^{\bf [1]}          $&$        2400~^{\bf [14]}         $&$                   1619 \pm   339  $\\
         NGC7674$~^{[a,b,c]}$&$  - $&$ 2.0 $&$                    448 \pm   110 $&$         718~^{\bf [8]}          $&$          50~^{\bf [29]}         $&$                   1095 \pm   248  $\\
             NGC7679$~^{[a]}$&$  - $&$ 1.0 $&$                    350 \pm    36 $&$         472~^{\bf [5]}          $&$         458~^{\bf [11]}         $&$                        -          $\\
         CGCG381-051$~^{[b]}$&$  - $&$ 2.0 $&$                          <    72 $&$           5~^{\bf [8]}          $&$                               -  $&$                    317 \pm   102  $\\
& & & & & & \\
\hline
& & & & & & \\
\multicolumn{7}{c}{Additional off-sample data}\\
& & & & & & \\
\hline
& & & & & & \\
                 NGC253$~^{}$&$-0.8$&$ 2.0 $&$                   1519 \pm   239 $&$                            -     $&$                               -  $&$                        -          $\\
                 NGC449$~^{}$&$ 1.7$&$ 2.0 $&$                               -  $&$         330~^{\bf [1]}          $&$          13~^{\bf [13]}         $&$                        -          $\\
                 NGC591$~^{}$&$ 1.4$&$ 2.0 $&$                               -  $&$        1780~^{\bf [3]}          $&$          18~^{\bf [11]}         $&$                        -          $\\
                 NGC613$~^{}$&$ 1.3$&$ 2.0 $&$                               -  $&$                            -     $&$                               -  $&$                        -          $\\
                  IC184$~^{}$&$ 1.4$&$ 2.0 $&$                               -  $&$                            -     $&$                               -  $&$                        -          $\\
                NGC1052$~^{}$&$ 2.1$&$ 2.0 $&$                               -  $&$                            -     $&$         112~^{\bf [36]}         $&$                        -          $\\
                NGC1106$~^{}$&$ 0.9$&$ 2.0 $&$                               -  $&$                            -     $&$                               -  $&$                        -          $\\
                MRK1066$~^{}$&$ 1.5$&$ 2.0 $&$                               -  $&$        5140~^{\bf [3]}          $&$          36~^{\bf [13]}         $&$                        -          $\\
         IRAS03355+0104$~^{}$&$ 2.7$&$ 2.0 $&$                               -  $&$          76~^{\bf [8]}          $&$                               -  $&$                        -          $\\
                  IC342$~^{}$&$-2.0$&$ 2.0 $&$                               -  $&$          34~^{\bf [59]}         $&$                               -  $&$                        -          $\\
                UGC3255$~^{}$&$ 1.2$&$ 2.0 $&$                               -  $&$                            -     $&$                               -  $&$                        -          $\\
                VIIZw73$~^{}$&$ 2.2$&$ 2.0 $&$                               -  $&$          74~^{\bf [8]}          $&$                               -  $&$                        -          $\\
                NGC2273$~^{}$&$ 0.8$&$ 2.0 $&$                    187 \pm    56 $&$         330~^{\bf [3]}          $&$          69~^{\bf [18]}         $&$                   1014 \pm   204  $\\
                  MRK78$~^{}$&$ 1.5$&$ 2.0 $&$                    792 \pm    82 $&$         653~^{\bf [60]}         $&$                               -  $&$                    422 \pm    91  $\\
                MRK1210$~^{}$&$ 1.9$&$ 2.0 $&$                    209 \pm    26 $&$         285~^{\bf [18]}         $&$         840~^{\bf [61]}         $&$                   1244 \pm   260  $\\
         2MASXJ08362280$~^{}$&$ 3.4$&$ 2.0 $&$                               -  $&$                            -     $&$                               -  $&$                        -          $\\
                NGC2979$~^{}$&$ 2.1$&$ 2.0 $&$                               -  $&$          11~^{\bf [18]}         $&$                               -  $&$                        -          $\\
                 IC2560$~^{}$&$ 2.0$&$ 2.0 $&$                    558 \pm    85 $&$         125~^{\bf [18]}         $&$          39~^{\bf [62]}         $&$                    469 \pm   115  $\\
                  MRK34$~^{}$&$ 3.0$&$ 2.0 $&$                    626 \pm    76 $&$         520~^{\bf [1]}          $&$          35~^{\bf [11]}         $&$                    244 \pm    67  $\\
                NGC3735$~^{}$&$ 1.3$&$ 2.0 $&$                               -  $&$         330~^{\bf [59]}         $&$                               -  $&$                        -          $\\
                NGC4258$~^{}$&$ 1.9$&$ 1.9 $&$                     76 \pm    13 $&$         100~^{\bf [63]}         $&$         837~^{\bf [16]}         $&$                    576 \pm   132  $\\
              ESO269-12$~^{}$&$ 3.0$&$ 2.0 $&$                               -  $&$           6~^{\bf [18]}         $&$                               -  $&$                        -          $\\
                NGC4945$~^{}$&$ 1.7$&$ 2.0 $&$                    320 \pm    49 $&$                            -     $&$        1300~^{\bf [64]}         $&$                        -          $\\
                NGC5495$~^{}$&$ 2.3$&$ 2.0 $&$                               -  $&$                            -     $&$                               -  $&$                        -          $\\
               Circinus$~^{}$&$ 1.3$&$ 2.0 $&$                   8599 \pm  1231 $&$          83~^{\bf [65]}         $&$        1400~^{\bf [66]}         $&$                  39011 \pm  7803  $\\
                NGC5643$~^{}$&$ 1.4$&$ 2.0 $&$                    940 \pm   142 $&$         800~^{\bf [1]}          $&$          84~^{\bf [18]}         $&$                        -          $\\
                NGC5728$~^{}$&$ 1.9$&$ 1.9 $&$                   1162 \pm   117 $&$         115~^{\bf [18]}         $&$         133~^{\bf [6]}          $&$                    220 \pm    44  $\\
                NGC5793$~^{}$&$ 2.0$&$ 2.0 $&$                               -  $&$                            -     $&$          13~^{\bf [11]}         $&$                        -          $\\
                NGC6240$~^{}$&$ 1.6$&$ 2.0 $&$                    236 \pm    35 $&$         202~^{\bf [5]}          $&$          91~^{\bf [36]}         $&$                    483 \pm    96  $\\
                NGC6264$~^{}$&$ 3.1$&$ 2.0 $&$                               -  $&$        3200~^{\bf [67]}         $&$                               -  $&$                        -          $\\
                NGC6323$~^{}$&$ 2.7$&$ 2.0 $&$                               -  $&$                            -     $&$                               -  $&$                        -          $\\
                NGC6300$~^{}$&$ 0.5$&$ 2.0 $&$                    304 \pm    33 $&$         140~^{\bf [68]}         $&$         860~^{\bf [69]}         $&$                    887 \pm   177  $\\
            Eso103-G035$~^{}$&$ 2.6$&$ 2.0 $&$                               -  $&$          43~^{\bf [18]}         $&$         907~^{\bf [70]}         $&$                        -          $\\
         IRAS19370-0131$~^{}$&$ 2.2$&$ 2.0 $&$                               -  $&$                            -     $&$                               -  $&$                        -          $\\
                NGC6926$~^{}$&$ 2.7$&$ 2.0 $&$                     45 \pm     7 $&$         241~^{\bf [5]}          $&$                               -  $&$                     64 \pm    13  $\\
             AM2158-380$~^{}$&$ 2.7$&$ 2.0 $&$                               -  $&$                            -     $&$                               -  $&$                        -          $\\
                NGC7479$~^{}$&$ 1.2$&$ 2.0 $&$                          <   136 $&$                            -     $&$         112~^{\bf [71]}         $&$                        -          $\\

\end{supertabular}
}
\end{document}